\documentclass[a4paper,11pt]{article}

\usepackage{jheppub}
\usepackage[T1]{fontenc}
\usepackage[utf8]{inputenc}
\usepackage{mathrsfs} 
\usepackage{multirow}
\usepackage{hyperref}
\usepackage{slashed}
\usepackage{color}
\usepackage{graphicx,subcaption}

\newcommand\mrm[1]{\mathrm{#1}}
\newcommand\MeV{\ \mathrm{MeV}}
\newcommand\GeV{\ \mathrm{GeV}}

\newcommand\TeV{\ \mathrm{TeV}}

\newcommand\ns{\ \mathrm{ns}}

\newcommand\mSigmaN{m_{h_2}}
\newcommand\mSigmaC{m_{h^+}}
\newcommand\SigmaN{{h_2}}

\newcommand\HN{{h_1}}

\newcommand\lambdaSigma{{b_4}}
\newcommand\lambdaSigmaH{{a_2}}
\newcommand\aSigmaH{{a_1}}

\DeclareMathOperator{\Tr}{Tr}

\newcommand\rlim{0.05}
\newcommand\rconf{95}
\newcommand\massBound{230}
\newcommand\colliderReachBound{330}
\newcommand\disappearingTrackBound{250}
\newcommand\problemCLs{84} 

\title{\boldmath Two-Step Electroweak Symmetry-Breaking: Theory Meets Experiment}

\author[a]{Nicole~F.~Bell,}
\author[a]{Matthew~J.~Dolan,}
\author[a,1]{Leon~S.~Friedrich,\note{Corresponding author.}}
\author[b,c,d]{Michael~J.~Ramsey-Musolf,}
\author[a]{and Raymond~R.~Volkas}
\affiliation[a]{ARC Centre of Excellence for Particle Physics at the Terascale,
  School of Physics, The University of Melbourne, Victoria 3010, Australia}
  \affiliation[b]{Amherst Center for Fundamental Interactions, Department of
  Physics, University of Massachusetts Amherst, Amherst, MA 01003, USA and California Institute of Technology, Pasadena, CA 91135 USA}
 \affiliation[c]{Tsung-Dao Lee Institute and  School of Physics and Astronomy, Shanghai Jiao Tong University, 800 Dongchuan Road, Shanghai, 200240 China}
\affiliation[d]{Kellogg Radiation Laboratory, California Institute of Technology, Pasadena, CA 91125 USA}
\emailAdd{n.bell@unimelb.edu.au} \emailAdd{matthew.dolan@unimelb.edu.au}
\emailAdd{leon.friedrich@unimelb.edu.au}
\emailAdd{mjrm@physics.umass.edu}
\emailAdd{raymondv@unimelb.edu.au}

\begin{document}

\hfill{ACFI-T20-01}

\abstract{We study the phenomenology of a hypercharge-zero SU(2) triplet
  scalar whose existence is motivated by two-step electroweak symmetry-breaking. 
  We consider both the possibility that the triplets are stable and
  contribute to the dark matter density, or that they decay via mixing with the
  standard model Higgs boson. The former is constrained by disappearing charged track
  searches at the LHC and by dark matter direct detection experiments, while the
  latter is constrained by existing multilepton collider searches. We find that
  a two-step electroweak phase transition involving a stable triplet with a negative quadratic term is ruled out by direct
  detection searches, while an unstable triplet with a mass less than $\massBound\GeV$ is excluded at $\rconf\%$ confidence level.}
\maketitle

\section{Introduction}

The origin of the baryon asymmetry of the universe is a major open problem in
particle physics and cosmology. Successful baryogenesis mechanisms require
extensions to the standard model (SM), as it has neither enough charge-parity
(CP) violation nor does it provide the necessary out-of-equilibrium conditions.
Electroweak baryogenesis provides one possible solution, and is particularly
attractive as its association with new electroweak scale physics means it is
testable experimentally via collider searches\cite{Ramsey-Musolf:2019lsf} and electric dipole
moment (EDM) measurements (for a review, see, {\it e.g.} \cite{Morrissey:2012db}).

There has been recent interest in the possibility of multi-step electroweak
phase
transitions~\cite{StepInto,ColorTwoStepPrec,MorriseyTwoStep,TwoStep,ColorTwoStep,TripletGravWaves}.
In such scenarios the electroweak phase transition consists of multiple
transitions, where initially an exotic scalar charged under SU(2) gains a vacuum
expectation value (VEV) before a second transition to the SM Higgs phase takes
place. This scenario is attractive because the extended scalar sector has enough
freedom to support a strongly first order transition, and the new CP violating
interactions can be partially hidden in the new scalar sector in order to avoid
tight EDM constraints. Two step phase transitions have been examined for a range
of extended scalar sectors, including SU(2) triplet scalar extensions~\cite{StepInto,TripletGravWaves}, two Higgs doublet
models~\cite{MorriseyTwoStep}
and coloured scalar extensions~\cite{ColorTwoStepPrec,ColorTwoStep}. Two-step transitions have also been studied in the context of scalar sector extensions containing real or complex singlets~\cite{Profumo:2007wc, Espinosa:2011ax, Curtin:2014jma, Jiang:2015cwa, Kurup:2017dzf, Chiang:2017nmu, Bell:2019mbn}. In these scenarios, electroweak symmetry breaking occurs only once -- during the final transition to the present Higgs phase.

The simplest\footnote{Simplest in the sense that it has the fewest additional
  physical particles, and the fewest new parameters present without imposing
  additional symmetries.} model that can feature the desired two step
electroweak symmetry breaking transition is the real SU(2) triplet scalar
$\Sigma\sim(1,3,0)$ extension to the SM (the $\Sigma$SM).
Such an electroweak scale triplet may arise from the breaking of a high-scale GUT, e.g., the {$\bold{210}$} of SO(10)~\cite{SO10SUSYTriplet}.
The phase transition
structure of the $\Sigma$SM has been examined by
refs.~\cite{StepInto,tripletLattice,TripletGravWaves}.
While~\cite{StepInto,tripletLattice} focused on phase transitions rather than collider physics,~\cite{TripletGravWaves} has studied the impact on collider phenomenology in more detail. However, they consider a dimension-5 effective operator involving the triplet that significantly modifies the phenomenology relative to the minimal triplet model that we study.
The general
phenomenology of minimal hypercharge-zero SU(2) triplet scalar extensions has
been studied
extensively~\cite{MJRMTripletPheno,LHCTripletPheno2013,planckTriplet,LHCTripletPheno},
with a significant focus on the prospects of having the neutral component of the
triplet be stable and thus provide some or all of the dark matter (DM)
density~\cite{StrumiaMinimalDM,StrumiaSommerfeld,StrumiaCosmic,MultipletEWPTDM,TripletDM1,LHCTripDM,TripDMFootprint}.

Ref.~\cite{MJRMTripletPheno} examines the prospect of constraining triplet scalars via measurements of the Higgs diphoton decay rate, disappearing track searches, and
collider production searches. However, as ref.~\cite{MJRMTripletPheno} was
published prior to first collisions at the LHC, no lower bounds on the triplet mass were set beyond those following from searches at the LEP collider. The more recent studies~\cite{planckTriplet,LHCTripletPheno} consider
corrections to SM Higgs production rates and decay processes, and do not obtain a lower bound on the triplet arising from the production and decay of the triplets at the LHC.
In the scenario where the neutral component of the triplet is stable, existing DM direct detection constraints severely restrict the size of the triplet's coupling to the SM Higgs.  
However, in order for the neutral component of the triplet make up a significant fraction of the DM density it is required to have a mass $\sim 2\TeV$. In contrast, acquiring a multi-step electroweak phase transition requires the mass to be electroweak-scale $\lesssim 1\TeV$. Thus the parameter-space relevant to multi-step phase transitions will only ever result in the triplet contributing a small fraction of the DM density and is generally not thoroughly explored in triplet scalar DM studies.

We extend the previous examinations of SU(2) triplet scalar phenomenology in a number of ways. 
Firstly, we show that if the neutral triplet is
stable or very long lived, then existing disappearing track searches constrain
the mass of the triplet to be larger than $\sim\disappearingTrackBound\GeV$. Secondly, we examine the
scenario where the neutral component of the triplet is both stable and 
hypothetically constitutes a portion of the dark
matter. We show that the parameter-space favourable for a multi-step
electroweak phase transition is ruled out by dark matter direct detection
experiments. Finally, we demonstrate that if the triplet is unstable, existing
LHC multilepton searches place a lower bound on its mass of around
$\massBound\GeV$. Utilising multilepton searches to constrain triplets has previously
been examined by refs.~\cite{TMSSM}~and~\cite{multileptonSeesaw,LeftRightSeesaw,SeesawCheckmate} in the context
of a triplet extended super-symmetric standard model, and a type-II
seesaw model, respectively. 

The above dark matter direct detection constraint implies that the neutral member of the
triplet must be allowed to decay if it is to be relevant for 2-step EWSB. The stability
of the neutral triplet in the $\Sigma$SM arises from the imposition of a $\Sigma
\rightarrow - \Sigma$ discrete $\mathbb{Z}_2$ symmetry on the model. This symmetry can be
broken explicitly by a term in the Lagrangian, so that the neutral triplet can
decay and the dark matter constraints are avoided. However, the collider production
constraints remain relevant. The advent of additional LHC data will increase the reach in both mass and coupling, thereby providing a powerful probe of this scenario.


\section{Model}
\label{sec:scalarPotential}
We extend the Standard Model by adding a real scalar field $\Sigma$ transforming as $(1,3,0)$ under the $SU(3)\times SU(2)\times U(1)_Y$ SM gauge group.  We consider the most
general renormalisable scalar potential,
\begin{equation}
  \begin{aligned}
  V_0(H,\Sigma) \ =& \  -\mu_H^2 H^\dagger H\  -\  \frac{1}{2}\mu_\Sigma^2
  \mathrm{Tr}(\Sigma^2)\ +\ \lambda_H (H^\dagger H)^2\ +\  \frac{1}{4}
  \lambdaSigma [\mathrm{Tr}(\Sigma^2)]^2 \\ &  \ +\ \frac{1}{\sqrt{2}} \aSigmaH H^\dagger \Sigma H \  + \ \frac{1}{2} \lambdaSigmaH \mathrm{Tr}(\Sigma^2) H^\dagger H \,, \label{eq:V0} 
\end{aligned}
\end{equation}
where $H$ is the SM scalar Higgs doublet, and we use the notation
  \begin{equation}
    \Sigma = \begin{bmatrix} \frac{1}{\sqrt{2}}  \left( \Sigma^0 + v_\Sigma \right) & \Sigma^+  \\  \Sigma^- & -\frac{1}{\sqrt{2}}   \left( \Sigma^0  +v_\Sigma \right)  \end{bmatrix} \, , \quad H = \begin{bmatrix} H^+ \\ \frac{1}{\sqrt{2}} (v_H + H^0 + i A^0)\end{bmatrix}\, .
  \end{equation}
  For real triplets, terms in the potential proportional to $\Tr \left(  \Sigma^4 \right)$  and $H^\dagger \Sigma^2 H$ can be absorbed into the $[\Tr (\Sigma^2 )]^2$ and $\Tr (\Sigma^2 ) H^\dagger H$ terms and simply redefine $\lambdaSigma$ and $\lambdaSigmaH$. We only consider negative quadratic coefficients for the triplet and the Higgs doublet.
To ensure that the potential is bounded from below we require
\begin{equation}
  \label{eq:BFB}
 \lambda_H>0,\ \lambdaSigma>0 , \  \lambdaSigmaH \geq - 2 \sqrt{\lambda_H \lambdaSigma}\,.
\end{equation}
Additionally, the vacuum at zero temperature must approximate the SM Higgs-phase within errors, so that
\begin{equation}
  m_{H} \simeq 125\GeV,\ v_H \simeq 246\GeV.
\end{equation}
The VEV of the triplet $v_\Sigma$ is constrained by precision electroweak
measurements as it contributes to the $\rho$ parameter. At tree level the
correction to the $\rho$ parameter is
\begin{equation}
 \delta \rho = \rho - 1 =  \frac{4 v_\Sigma^2}{ v_H^2} .
\end{equation}
The current measurement of $\rho = 1.00039\pm 0.00019$~\cite{PDG2018} requires $v_\Sigma \lesssim 3\GeV$.

We  consider two scenarios: a model where we impose a $\Sigma
\rightarrow -\Sigma$ discrete $\mathbb{Z}_2$ symmetry on the theory, which eliminates the
$\aSigmaH$ coupling, and a model with no such symmetry, where
$\aSigmaH \neq 0$. In the remainder of this section we discuss
the notation and selection of parameters in each scenario before moving on to
discuss perturbativity constraints, electroweak phase transition requirements, and corrections to the SM Higgs diphoton rate.

\subsection{$\mathbb{Z}_2$ symmetric model}
\label{sec:symmMod}
With the $\Sigma \rightarrow -\Sigma$ symmetry imposed on
the theory, the potential has four permissible types of extrema~\cite{MJRMTripletPheno}:
\begin{enumerate}
  \item $v^2_H = 0$, $v^2_\Sigma = 0$
  \item $v^2_H = 0$, $v^2_\Sigma = \frac{\mu_\Sigma^2}{\lambdaSigma}$
  \item $v_H^2 = \frac{\mu_H^2}{\lambda_H}$, $v^2_\Sigma = 0$
  \item $v_H^2 = \frac{4 \lambdaSigma \mu_H^2 - 2 \lambdaSigmaH
      \mu_\Sigma^2 }{ 4 \lambda_H \lambdaSigma - \lambdaSigmaH^2 } $,
      $v^2_\Sigma = \frac{4 \lambda_H \mu_\Sigma^2 - 2 \lambdaSigmaH
      \mu_H^2 }{ 4 \lambda_H \lambdaSigma - \lambdaSigmaH^2 }$
\end{enumerate}
Only the latter two can yield SM-like minima, since $v_H \neq 0$. However, the
fourth possibility results in a physical charged scalar that is massless at tree-level. This is due to
the fact that the $\mathbb{Z}_2$ symmetric potential features only
$\mathrm{Tr}(\Sigma^2)$ terms, leading to an accidental $SO(3)$ global symmetry
which rotates the components of $\Sigma$ amongst themselves but under which $H$ is a singlet. This symmetry is
spontaneously broken when the triplet gains a VEV, yielding a charged pseudo-Goldstone
scalar boson.

Therefore we focus on the scenario where the zero temperature potential
has a global minimum of the third type. This extremum is a local minimum when the parameters
satisfy
\begin{equation}
  \label{eq:higsMinCond}
  \frac{\mu_H^2}{\lambda_H} = v_H^2 > 2 \frac{\mu_\Sigma^2}{\lambdaSigmaH}\, ,
\end{equation}
and is the global minimum when
\begin{equation}
  \label{eq:stability}
 \frac{\mu_H^4}{\lambda_H}  > \frac{\mu_\Sigma^4 }{\lambdaSigma} \, .
\end{equation}
The Higgs couplings then take
their SM values, $\lambda_H=\frac{m_{H}^2}{2 v_H^2}$ and $\mu_H^2 =
\frac{m_{H}^2}{2}$. The potential has three free parameters $\lambdaSigmaH$,
$\mu_\Sigma^2$, and $\lambdaSigma$. We swap $\lambdaSigmaH$ for the triplet mass using the relation
\begin{equation}
  \label{eq:tripletMass}
  m^2_{\Sigma^0} = -\mu_\Sigma^2 + \frac{1}{2} \lambdaSigmaH v_H^2.
\end{equation}

The form of the $\mathbb{Z}_2$-symmetric potential has the triplet components being degenerate at tree-level.
However, radiative corrections lead to a small mass splitting between the neutral
and charged components~\cite{StrumiaMinimalDM},
\begin{equation}
  \label{eq:splitting}
  \Delta m_\Sigma = m_{\Sigma^+} - m_{\Sigma^0} = \frac{\alpha_2 m_{\Sigma^0} }{4 \pi}\left[ f\left( \frac{m_W}{m_{\Sigma^0}} \right)  - c_W^2  f\left(\frac{m_Z}{m_{\Sigma^0}} \right)   \right] > 0,
\end{equation}
where $c_W$ is the cosine of the weak mixing angle and,
\begin{equation}
  f(r) = -\frac{r}{4} \left[ 2 r^3 \ln r + \left(r^2 - 4  \right)^{3/2} \ln \left(  \frac{r^2 - 2 - r \sqrt{r^2 - 4} }{2}\right) \right].
\end{equation}
The mass splitting decreases with increasing triplet mass, and in the limit 
$\frac{m_{\Sigma^0}}{m_Z} \gg 1$ the mass splitting approaches
$\Delta m_\Sigma = 166\MeV$. While the neutral component remains stable, this
small splitting allows the charged component to decay via an off shell $W^\pm$
into the neutral component and either a low energy pion or a light charged lepton
and neutrino. The widths of the associated decays are given
by~\cite{StrumiaMinimalDM,neutronDecay}
\begin{subequations}
  \label{eq:lifetime}
\begin{align}
  \Sigma^+ \rightarrow \Sigma^0 \pi^+ \ & :\ \Gamma_\pi = \frac{2 G_{\mrm{F}}^2 \left\lvert  V_{u d}\right\rvert^2 {\Delta m_\Sigma}^3 f_\pi^2 }{\pi} \sqrt{1 - \frac{m_\pi^2}{{\Delta m_\Sigma}^2}} \, , \\
  \Sigma^+ \rightarrow \Sigma^0 e^+ \nu_e  \ & :\ \Gamma_e = \frac{2 G_{\mrm{F}}^2 {\Delta m_\Sigma}^5}{15 \pi^3} \, ,\\ 
  \Sigma^+ \rightarrow \Sigma^0 \mu^+ \nu_\mu  \ & :\   \Gamma_{\mu} = K\left( \frac{m_\mu}{{\Delta m_\Sigma}} \right) \Gamma_e ,
\end{align}
\end{subequations}
where $f_\pi \simeq 131\MeV$ and
\begin{equation}
  K(x) = \frac{15}{2} x^4 \log\frac{1 + \sqrt{1 - x^2}}{x} - \frac{1}{2}\sqrt{1-x^2}\left( 8 x^4 + 9 x^2 - 2 \right).
  \end{equation}

\subsection{$\mathbb{Z}_2$ broken model}

Turning on the $\mathbb{Z}_2$-breaking $\aSigmaH$ term changes the
results of the previous subsection. In particular, for the third type of extremum
the triplet gains a small 
induced VEV from the $H^2 \Sigma$ term, with the potential now minimised by
\begin{subequations}
  \label{eq:vevs}
\begin{align}
  v_\Sigma &=
  \frac{\aSigmaH v_H^2}{- 4 \mu_\Sigma^2 + 2 v_H^2 \lambdaSigmaH + 4 v_\Sigma^2 \lambdaSigma} \nonumber
  \\
   & \simeq \frac{\aSigmaH v_H^2}{- 4 \mu_\Sigma^2 + 2 v_H^2 \lambdaSigmaH} = \frac{\aSigmaH v_H^2}{4 m_{\Sigma^0}^2} \\
  v_H^2 & = \frac{\mu_H^2}{\lambda_H}  + \frac{\aSigmaH v_\Sigma -  \lambdaSigmaH v_\Sigma^2 }{2 \lambda_H} \simeq \frac{\mu_H^2}{\lambda_H},
\end{align}
\end{subequations}
where $m_{\Sigma^0}$ is the mass of the triplet in the $\mathbb{Z}_2$ symmetric
case, eq.~\eqref{eq:tripletMass}, and the approximations hold when the triplet
VEV is small. 

Additionally the $\aSigmaH$ term in the potential and the
triplet's non-zero VEV 
result in new mass terms leading to mixing between the neutral component of the
triplet and SM Higgs,
\begin{equation}
\begin{aligned}
  \mathcal{L} & \supset
  \frac{1}{2}\begin{pmatrix}
    H^0 & \Sigma^0 
  \end{pmatrix}
  \mathcal{M}_{N}
  \begin{pmatrix}
    H^0 \\ \Sigma^0 
  \end{pmatrix},
  \\ & = \frac{1}{2}\begin{pmatrix}
    h_1 & h_2 
  \end{pmatrix}
  \begin{pmatrix}
    m_{h_1}^2 & 0 \\ 0 & m_{h_2}^2
  \end{pmatrix}
  \begin{pmatrix}
    h_1 \\ h_2 
  \end{pmatrix}, 
\end{aligned}
\end{equation}
where we have introduced the neutral scalar mass matrix,
\begin{align}
  \mathcal{M}_N &= 
  \begin{pmatrix}
    2 \lambda_H v_H^2  \  &\  \lambdaSigmaH v_H v_\Sigma - \frac{1}{2} \aSigmaH v_H 
    \\
   \lambdaSigmaH v_H v_\Sigma - \frac{1}{2}\aSigmaH v_H\ &\ - \mu_\Sigma^2 + \frac{1}{2} \lambdaSigmaH v_H^2 + 3 \lambdaSigma v_\Sigma^2   
 \end{pmatrix},
\end{align}
and the mass basis,
\begin{align}
  \begin{pmatrix}
    h_1 \\ h_2 
  \end{pmatrix}
  &= 
  \begin{pmatrix}
    \cos \theta_N &\   - \sin \theta_N \\
    \sin \theta_N &\  \cos \theta_N
  \end{pmatrix} 
  \begin{pmatrix}
    H^0 \\ \Sigma^0 
  \end{pmatrix}.
\end{align}
The neutral scalar mixing angle $\theta_N$ is defined such that $h_1$ is the
particle that consists primarily of $H^0$. As we require $v_\Sigma \lesssim 3
\GeV$, and as the off-diagonal term is directly proportional to $v_\Sigma$, the
mixing term is necessarily small. Hence, unless the scalars are
nearly degenerate the mixing angle will also be small. It is then sufficient to use
the SM values for $\mu_H$ and $\lambda_H$ in order to produce a SM-like Higgs
with $m_{h_1} \approx 125\GeV$ and $v_H \approx 246 \GeV$. The potential then
has four free parameters: $\mu_\Sigma^2$, $\lambdaSigmaH$,
$\aSigmaH$, and $\lambdaSigma$. We will fix $\lambdaSigmaH$
and $\aSigmaH$ by requiring that we get values for $m_{h_2}$
and $v_\Sigma$, as given by diagonalising $\mathcal{M}_N$ and solving
eq.~\eqref{eq:vevs}, respectively. 

There will also be mixing in the charged scalar sector,
\begin{align}
  \mathcal{L} \  & \supset\ 
  \begin{pmatrix}
    H^- & \Sigma^- 
  \end{pmatrix}
          \mathcal{M}_C
  \begin{pmatrix}
    H^+ \\ \Sigma^+ 
  \end{pmatrix} 
  \  =\  \begin{pmatrix}
    G^- & h^- 
  \end{pmatrix}
  \begin{pmatrix}
    0 & 0 \\ 0 & m_{h^+}^2
  \end{pmatrix}
  \begin{pmatrix}
    G^+ \\ h^+ 
  \end{pmatrix}  , 
  \end{align}
  where
  \begin{subequations}
  \begin{align}
  \mathcal{M}_C &= 
   \frac{\aSigmaH}{4}\begin{pmatrix} 4 v_\Sigma & 2 v_H \\ 2 v_H & v_H^2/v_\Sigma \end{pmatrix} , \\
  \begin{pmatrix}
    G^+ \\ h^+ 
  \end{pmatrix}
  & = 
  \begin{pmatrix}
    \cos \theta_C &\   - \sin \theta_C \\
    \sin \theta_C &\  \cos \theta_C
  \end{pmatrix} 
  \begin{pmatrix}
    H^+ \\ \Sigma^+ 
  \end{pmatrix}\, , \quad
\sin{\theta_C} = \frac{v_\Sigma}{\sqrt{v_\Sigma^2 + \frac{1}{4} v_H^2 }}. 
\end{align}
\end{subequations}
The field $G^+$ is the massless charged unphysical Goldstone boson and $h^+$ is a physical
charged scalar that consists primarily of the charged triplet component
$\Sigma^+$.

In the limit $v_\Sigma \rightarrow 0$ we re-obtain the $\mathbb{Z}_2$ symmetric model and the masses of the scalars approach the values they would have had in the absence of mixing, $m_{h_1}\rightarrow m_{H}$, $m_{h_2}\rightarrow m_{\Sigma^0}$, and $m_{h^+} \rightarrow m_{\Sigma^+}$. For simplicity, we will use the notation of the $\mathbb{Z}_2$ broken model to identify particles and masses throughout the remainder of the paper, even if there is no mixing.
Note that this limiting behaviour means that the radiative mass splitting discussed in the previous subsection will
become important for very small $v_\Sigma$.
However, unless $v_\Sigma \lesssim 10^{-3} \GeV
$~\cite{MJRMTripletPheno} the charged scalar will primarily decay via its mixing
with the charged Goldstone boson into pairs of fermions or $W^\pm Z^{(*)}$, and
not via the decays discussed in the previous section. Hence, unless $v_\Sigma$ is
very small the decays will not be sensitive to the radiative mass splitting. We discuss
the unstable triplet decays in detail in section~\ref{sec:unstabledecays}.

\subsection{Perturbative Unitarity and Perturbativity}

Requiring that our couplings satisfy perturbative unitarity, i.e.\ that the tree-level high energy $2\rightarrow 2$ scattering amplitudes remain unitary, leads to the constraints~\cite{planckTriplet,LHCTripletPheno},
\begin{subequations}
\begin{align}
  &\lvert \lambdaSigmaH \rvert \leq 8 \pi \, , \\
  &\lvert  \lambda_H \rvert,\, \lvert \lambdaSigma  \rvert \leq 4 \pi \, , \\
  &\lvert 6 \lambda_H + 5 \lambdaSigma \pm \sqrt{(6 \lambda_H - 5 \lambdaSigma)^2 + 12 \lambdaSigmaH} \rvert \leq 16 \pi \, ,
\end{align}
\end{subequations}
where we have utilised the unitarity constraint with $\left\lvert \mrm{Re}(a_0)
\right\rvert \leq \frac{1}{2}$. Combining these constraints with the
requirement that the potential be bounded from below, eq.~\eqref{eq:BFB}, the
constraints on the couplings become,
\begin{subequations}
\label{eq:scalarBounds}
\begin{align}
  &0 \leq \lambda_H \leq \frac{4}{3}\pi \, , \\
  &0 \leq \lambdaSigma \leq \frac{8}{5}\pi \, , \label{eq:lambdaSigmaBound} \\
  \label{eq:lambdaSigmaHBound}
  & \lvert  \lambdaSigmaH \lvert \leq \sqrt{10 \left( \lambda_H - \frac{4}{3} \pi \right) \left( \lambdaSigma - \frac{8}{5} \pi \right)} \lesssim 4.54 \pi \, .
\end{align}
\end{subequations} 

While well defined, the perturbative unitarity requirement is separate from the requirement that the scalar couplings be perturbative. The definition of perturbativity is somewhat subjective. One method of defining a perturbativity bound is via the renormalization group equations (RGEs). In the SM at one-loop level, the Higgs quartic coupling features a Laundau pole at high energy.  On the other hand, the two-loop RGEs instead have the quartic coupling approaching a fixed point $\lambda_H(\mu)\rightarrow \lambda_{H}^{\mathrm{FP}}\approx 12$~\cite{HiggsPerturbativity, HiggsPerturbativity2}. When $\lambda_H = \lambda_{H}^{\mathrm{FP}}$ the two-loop contributions to the RGEs cancel the one-loop terms, therefore the fixed point provides a value of the coupling at which perturbativity begins to break down. This same behaviour is present in the real triplet scalar extended standard model. Therefore, following refs.~\cite{HiggsPerturbativity, HiggsPerturbativity2,mjrmRunningCouplingsSinglets, mjrmRunningCouplings}, we impose the requirement,
\begin{equation}
\label{eq:perturbativity}
\lambda < \frac{\lambda^{\mathrm{FP}}}{3} \, , \quad \lambda \in \{\lambda_H, \lambdaSigma, \lambdaSigmaH\} \, ,
\end{equation}
where $\lambda^{\mathrm{FP}}$ is the fixed point of each of the scalar couplings. 
We utilise the \texttt{SARAH}~$4.14.3$~\cite{SARAH4} package, which has an implementation of the real triplet extension, to evaluate the two-loop RGEs. We find that for a wide range of initial conditions, the scalar couplings approach the fixed points,
\begin{equation}
\label{eq:fixedPoints}
\lambda_{H}^{\mathrm{FP}} \approx 12 \, , \quad b_{4}^{\mathrm{FP}} \approx 6 \, , \quad a_{2}^{\mathrm{FP}} \approx 23 \, .
\end{equation}
Thus, our perturbativity requirement is,
\begin{equation}
\label{eq:perturbativity2}
\lambda_{H} < 4 \, , \quad b_{4} < 2 \, , \quad a_{2} < 7.7 \, .
\end{equation}
With the exception of the Higgs quartic coupling, this perturbativity condition is significantly more restrictive than the perturbative unitarity requirement from eq.~\eqref{eq:scalarBounds}.

\subsection{RGEs and running constraints}

\begin{figure}
  \centering
    \includegraphics[width=0.95\textwidth]{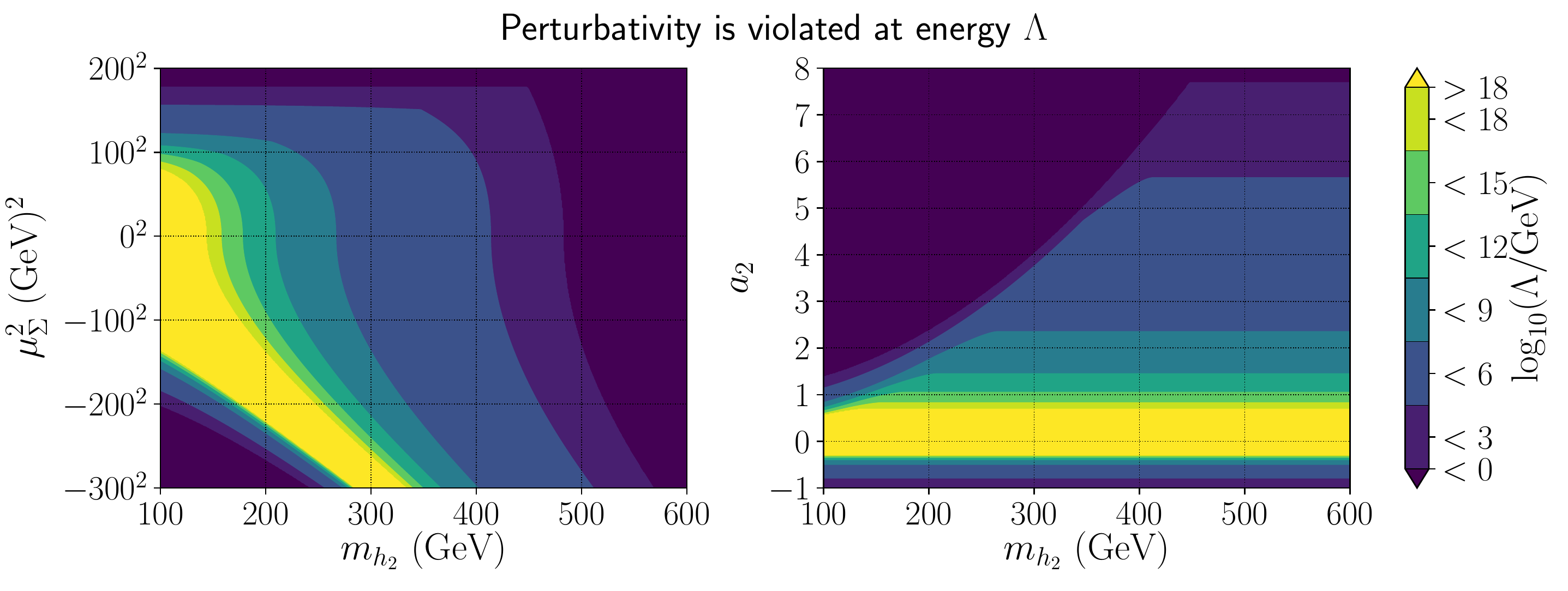}
    \caption{
  \label{fig:rge}
  Contour plots showing the energies at which the perturbativity and perturbative unitarity constraints are violated as a function of $\mSigmaN$ and either $\mu_\Sigma^2$ (left) or $\lambdaSigmaH$ (right). The darkest shade corresponds to the region of parameter-space where the conditions are not satisfied by the initial choice of parameters, with no running necessary. Conversely, the lightest shade corresponds to the region where the conditions are still satisfied after running the couplings up to very high energies. For each point we have set $v_\Sigma=0$, and $b_4$ to the minimal value allowed by eq.~\eqref{eq:stability}. This choice of $b_4$ was found to maximise the energy at which the conditions were first violated.
  }
\end{figure}

We also require that the perturbativity and perturbative unitarity conditions continue to be satisfied at higher energy scales, up to some cutoff energy $\Lambda$. In particular, if a set of parameters lies near the non-perturbative region and one uses the RGEs to run the couplings they may rapidly become non-perturbative even at relatively low energies ($ \sim 1\TeV$). The choice of cutoff energy significantly impacts the amount of parameter-space available. Figure~\ref{fig:rge} shows how the available parameter-space depends on the energy cutoff. Requiring that the perturbativity conditions are satisfied up to $\Lambda=10^6\GeV$ or higher removes a large chunk of the available parameter-space. We consider the requirement that the couplings continue to be perturbative up to at least $\Lambda = \mSigmaN, \mSigmaC \sim 1\TeV$ to be the bare-minimum requirement that we will impose for the remainder of the paper, though we will also consider more restrictive higher energy cut-offs. However, if there are additional light particles ($m < 10^6 \GeV$) that strongly couple to the SM Higgs or triplet, then they may significantly modify the running. Hence, even requiring perturbativity and perturbative unitarity only up to $\Lambda=10^6\GeV$ may be excessive if one expects such new physics.

\subsection{Phase transition requirements}

We study a model where, in the early universe, electroweak symmetry
breaking occurs via a transition from the electroweak symmetric minimum to a
minimum where the scalar triplet gains a VEV. A subsequent transition then
takes us to the regular SM-like Higgs phase at a lower temperature. Requiring
such a multi-step electroweak phase transition leads to constraints on the
scalar potential parameters. 

An important necessary condition is that the triplet should have a negative quadratic coefficient: $-\mu^2_\Sigma < 0$.
To see this, consider the opposite situation, that $-\mu^2_\Sigma > 0$, where we deal
with the $\mathbb{Z}_2$-symmetric model first. At finite temperature,
the tendency is for a quadratic coefficient to gain a positive contribution so that 
$-\mu^2_\Sigma = |\mu^2_{\Sigma}| \to |\mu^2_{\Sigma}| + a T^2$, where $a>0$.\footnote{If there is a large negative $\lambdaSigmaH$ coupling it is possible for the thermal term to be negative, leading to symmetry non-restoration. However, in our model this is incompatible with the requirement that the potential be bounded from below, eq.~\eqref{eq:BFB}.} In isolation, this effect
goes against our desire for $\Sigma$ to have a nonzero VEV at finite temperature and thus participate
in a two-step electroweak phase transition. The only way out is for a sufficiently large negative
quadratic coefficient to be induced from a negative $\lambdaSigmaH$ coupling such that the effective
quadratic coefficient is negative: $|\mu_\Sigma^2| + a T^2 + \frac{1}{2}\lambdaSigmaH v_H^2 < 0$.
But if this were the case, then at zero temperature the large and negative $\lambdaSigmaH$ would induce
a large triplet VEV, which is ruled out from the $\rho$-parameter bound. Thus the opposite choice of 
$-\mu^2_{\Sigma} < 0$ is the
only viable possibility, and we adopt it as a necessary though not sufficient condition to have
an acceptable two-step electroweak phase transition.\footnote{The feature of the
  third extremum that $v_\Sigma = 0$
at zero temperature then requires that the $\lambdaSigmaH v^2_H$ induced contribution be sufficiently
large so that the effective quadratic coefficient $-\mu_\Sigma^2 + \frac{1}{2}\lambdaSigmaH v_H^2$
is both positive and large enough to produce phenomenologically-viable triplet scalar masses. In this
situation, $\lambdaSigmaH$ must be positive.} This is consistent with the 
parameter space explored in previous multi-step phase transition
models~\cite{ColorTwoStepPrec,MorriseyTwoStep,TwoStep,ColorTwoStep},
particularly refs.~\cite{StepInto,tripletLattice,TripletGravWaves}. The $\mathbb{Z}_2$-broken case follows similarly, with
the only change being that the triplet gains a small induced VEV at zero temperature from the cubic $\aSigmaH$ term. Requiring that the VEV be small necessitates that $\aSigmaH$ is small, such that it has no significant impact on early universe phase transitions aside from breaking the $\mathbb{Z}_2$ symmetry.

A rigorous treatment of the finite temperature effective potential and early
universe phase transitions is non-trivial, with significant theoretical and
technical issues remaining to be addressed. In particular the typical phase
transition treatments are gauge dependent~\cite{MJRMgaugeDep}. However, even
one-loop gauge-independent treatments lead to results that differ from current
lattice simulations~\cite{MJRMgaugeDep,LatticeGravWave}. Accordingly, it is difficult to make
precise statements about the requirements that should be placed on the scalar
potential couplings to obtain the desired phase transition. Therefore, we simply
use the arguments presented and focus on triplets with negative quadratic terms
$-\mu_\Sigma^2<0$ and, as a consequence, positive Higgs couplings
$\lambdaSigmaH>0$. One potential caveat is that for models with further
extensions to the scalar sector, it is possible that some other particle (e.g.\
a scalar singlet) may have gained a VEV that acts to destabilise the triplet in
the early universe, or may have a VEV at zero-temperature acting to increase the
mass of the triplet~\cite{ColorTwoStepPrec}. This allows for the possibility
that $-\mu_\Sigma^2>0$ while still letting the triplet gain a VEV in the early
universe. Hence, we will also examine the parameter space where $-\mu_\Sigma^2$
takes on small positive values $- \mu_\Sigma^2 \sim (100 \GeV)^2 $, despite the
fact that such further extensions might significantly affect the phenomenology.

Combining the requirement that $\mu_\Sigma^2 > 0$ with the requirement that the scalar couplings satisfy perturbativity and perturbative unitarity then directly leads to an upper bound on the mass of the triplet. From figure~\ref{fig:rge}, we see that requiring perturbativity up to $\Lambda=1\TeV$ requires $\mSigmaN\lesssim 415 \GeV$. If we instead require perturbativity up to $10^6 \GeV$, this upper bound decreases to $\mSigmaN \lesssim 270\GeV$.

\subsection{Higgs Diphoton Rate}

In the SM the Higgs can decay into two photons via a fermion or $W^\pm$ loop.
The introduction of the triplet scalar will lead to a correction to the SM Higgs
diphoton rate via the addition of a new charged scalar loop. This correction is
proportional to $\lambdaSigmaH$ and decreases with increasing charged scalar mass.
However, in our scenario a larger mass necessarily means a larger
$\lambdaSigmaH$, and hence a precise measurement of the diphoton rate could
in principle be used to exclude triplets with negative quadratic coefficients
altogether. The SM diphoton rate is given by~\cite{diphotonSM}
\begin{equation}
\label{eq:SMdiphoton}
  \Gamma^{\mathrm{SM}}_{H \rightarrow \gamma \gamma}
  =
    \frac{\alpha^2 g_2^2}{1024 \pi^3} \frac{m_H^3}{m_W^2}
    \ \left\lvert \frac{4}{3} F_{1/2}\left(4 \frac{m_t^2}{m_H^2}\right) + F_1 \left(4 \frac{m_W^2}{m_H^2}\right) \right\rvert ^2 \, .
\end{equation}
Neglecting the small charged scalar mixing angle $\theta_C$, the triplet modifies the diphoton rate to~\cite{TwoStep},
\begin{equation}
\begin{aligned}
  \Gamma^{\mathrm{\Sigma SM}}_{h_1 \rightarrow \gamma \gamma}
  \approx & 
    \frac{\alpha^2 g_2^2}{1024 \pi^3} \frac{m_{h_1}^3}{m_W^2}
     \left\lvert\
      \frac{4}{3} \cos \theta_N
      F_{1/2}\left(4 \frac{m_t^2}{m_{h_1}^2}\right)
      \right. \\
      & \  +  \left( \cos \theta_N - \sin \theta_N \frac{4 v_\Sigma}{v_H}\right)
      F_1 \left(4 \frac{m_W^2}{m_{h_1}^2}\right) 
      \\
      &\   + \left.\left( \cos \theta_N \lambdaSigmaH - \sin \theta_N \lambdaSigma \frac{2 v_\Sigma }{v_H}  \right)
      \frac{v_H^2}{ 2 m_{h^+}^2}
      F_{0}\left(4 \frac{m_{h^+}^2}{ m_{h_1}^2}\right) \right\rvert ^2    \, ,
\end{aligned}
\end{equation}
where the loop functions are,
\begin{subequations}
  \begin{align}
    F_{0} (x) &= x (1 - x f(x)) \\
    F_{1/2} (x) &= - 2 x (1 + (1 - x)f(x)) \\
    F_{1} (x) &= 2 + 3 x (1 + (2-x)f(x))\\
    f(x) & =
           \begin{cases} 
             \mrm{arcsin}^2 (1/\sqrt{x}) & x\geq 1 \\
             - \frac{1}{4}\left( \ln\frac{1 + \sqrt{1-x}}{1-\sqrt{1-x}} - i \pi \right)^2 &  x < 1 \\
           \end{cases} \, .
  \end{align}
\end{subequations}
The $\mathbb{Z}_2$ symmetric result can be obtained by setting $\theta_N$ and $v_\Sigma$ to zero.
The scalar consisting primarily of the triplet can also decay into two photons, with rate given by
\begin{equation}
\begin{aligned}
  \Gamma^{\mathrm{\Sigma SM}}_{h_2 \rightarrow \gamma \gamma}
  \label{eq:TripDiphoton}
  \approx &
    \frac{\alpha^2 g_2^2}{1024 \pi^3} \frac{m_{h_2}^3}{m_W^2}
    \   \left\lvert
 \frac{4}{3} \sin \theta_N
    F_{1/2}\left(4 \frac{m_t^2}{m_{h_2}^2}\right)
     \right. \\
    &\ + 
    \left( \sin \theta_N - \cos \theta_N \frac{4 v_\Sigma}{v_H}\right)
    F_1 \left(4 \frac{m_W^2}{m_{h_2}^2}\right)
    \\
    &\ + \left.\left( \sin \theta_N \lambdaSigmaH + \cos \theta_N \lambdaSigma \frac{2 v_\Sigma }{v_H}  \right)
      \frac{v_H^2}{ 2 m_{h^+}^2}
    F_{0}\left(4 \frac{m_{h^+}^2}{m_{h_2}^2}\right) \right\rvert ^2 \, .
\end{aligned}
\end{equation}

\begin{figure}
  \centering
    \includegraphics[width=0.95\textwidth]{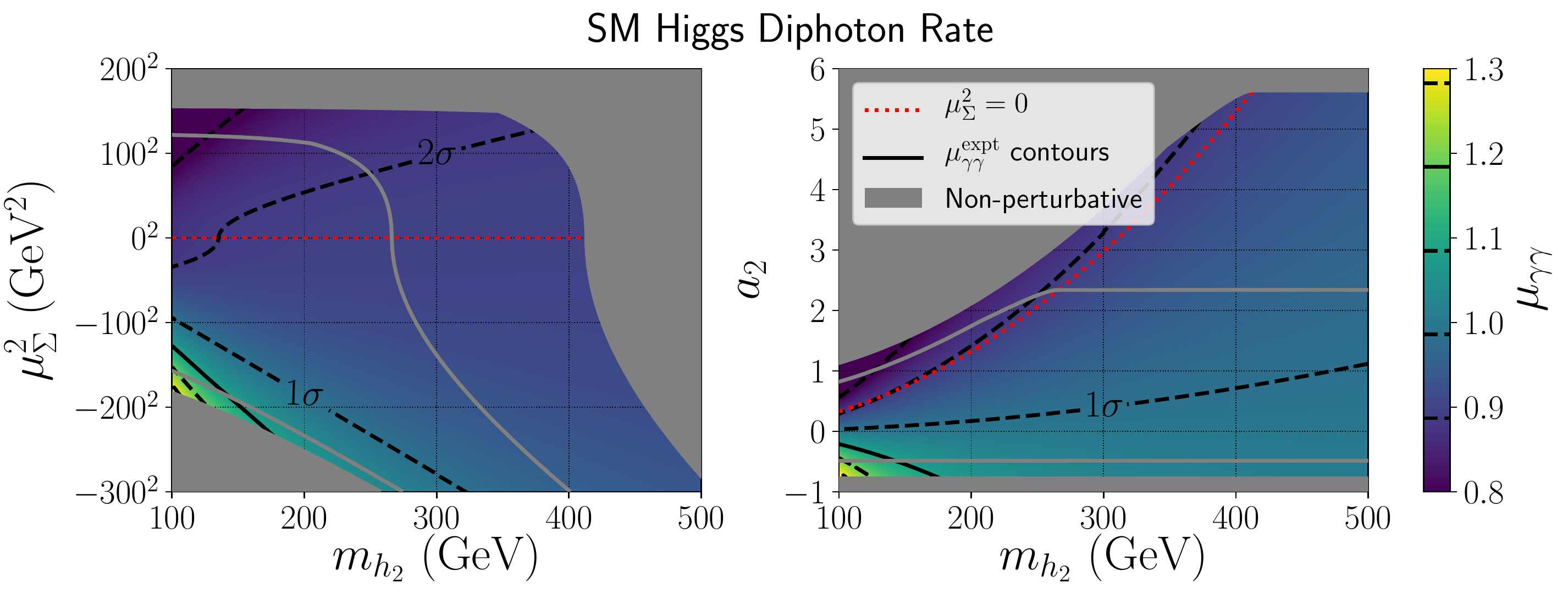}
    \caption{
  \label{fig:diphoton}
  SM Higgs diphoton rate as a function of the triplet-like neutral scalar mass $\mSigmaN$
  and either $\mu^2_\Sigma$ (left) or $\lambdaSigmaH$ (right), with $v_\Sigma=0$ and $\lambdaSigma=1$. The solid black line is the
  contour of the combined CMS and ATLAS diphoton rate measurement, and the
  dashed lines give the one-, two- and three-sigma contours.
  The solid grey region is the parameter-space where the scalar couplings become non-perturbative at energies $\Lambda < 1 \TeV$, and the solid grey line shows where this contour would be if the cutoff energy is increased to $\Lambda=10^6\GeV$. The red dotted line indicates the $\mu_\Sigma^2 = 0$ contour.}
\end{figure}

The signal strength of the SM Higgs to diphoton process is then given by
\begin{equation}
  \mu_{\gamma \gamma}
  =
    \frac{\Gamma^{\mathrm{\Sigma SM}}_{h_1 \rightarrow \gamma \gamma }}{\Gamma^{\mathrm{SM}}_{H \rightarrow \gamma \gamma}}  .
\end{equation}
We compare this with the most recent measurements by the
ATLAS~\cite{ATLASDiphoton} and CMS~\cite{CMSDiphoton} collaborations,
\begin{equation}
  \mu^{\mrm{ATLAS}}_{\gamma \gamma} = 0.99 \pm 0.14 \ , \quad  \mu^{\mrm{CMS}}_{\gamma \gamma} = 1.18^{+0.17}_{-0.14} \, .
  \end{equation}
We combine these measurements using a simple inverse variance weighted average.
Taking $0.14^2$ to be the variance of the CMS measurement this yields,
\begin{equation}
\mu^{\mathrm{expt}}_{\gamma \gamma} = 1.085 \pm 0.099 \, .
\end{equation}
Figure~\ref{fig:diphoton} shows the contour plots of the SM Higgs diphoton signal
strength as a function of $\mSigmaN$ and either $\mu_\Sigma^2$ or $\lambdaSigmaH$ in the $\mathbb{Z}_2$ symmetric case. The $\mathbb{Z}_2$ broken case only differs significantly near $m_{h_2} \approx m_{h_1}$, where the triplet-Higgs mixing angles are large.

The future prospects for measuring the SM Higgs diphoton signal strength at the
High-Luminosity LHC indicate an expected error of $\sim 10\%$ with $3\
\mrm{ab}^{-1}$ of data at $14\TeV$~\cite{prospectsATLAS,prospectsCMS}. Assuming
the measured value moves towards the SM prediction $\mu_{\gamma \gamma}=1$, this
enhanced accuracy will not result in constraints that are significantly more stringent than
the current ones, as the shift towards a SM value would offset the
decrease in error.

\section{Stable Triplet Phenomenology}
\label{sec:stablePheno}

\subsection{Disappearing Tracks}
\label{sec:dissaparingTracks}
As discussed in section~\ref{sec:symmMod}, in the $\mathbb{Z}_2$ symmetric model the small radiative mass splitting
allows the charged triplet component to decay via an off-shell $W^\pm$ into a neutral triplet component and a low energy
pion or lepton pair. As the triplet mass varies from $100\GeV$ to $1\TeV$, the
lifetime varies between $0.1$--$0.18\ns$ ($c\tau=3$--5~cm). Hence, as pointed
out by refs.~\cite{StrumiaMinimalDM,MJRMTripletPheno} charged triplets may
result in disappearing charged tracks at the LHC. Recent searches for
disappearing tracks produced by decaying charginos were performed by the
CMS~\cite{DissapearingTracksCMS} and ATLAS~\cite{DissapearingTracksATLAS}
collaborations using $36\ \mrm{fb}^{-1}$ of data.
The ATLAS disappearing track searches are more sensitive to small lifetimes than
the CMS searches. As the triplets will have small lifetimes, the ATLAS
searches provide the most severe constraints.
The ATLAS analysis provides a model-independent $95\%$
confidence upper bound on the visible cross section, alongside efficiency times acceptance data for  the production of charginos as a function of their lifetime and mass~\cite{DissapearingTracksATLAS,ATLASHEPDATA}. One of the production mechanisms considered in the ATLAS analysis is pair-production of charginos via charged or neutral current Drell-Yan processes, with cuts applied to the initial state
radiation jets and disappearing charged tracks.
Charged and neutral current Drell-Yan processes are also the dominant pair production processes for the charged triplets.
Hence, we directly take the chargino acceptance times efficiency data, linearly interpolate it, and apply it to the charged
triplet production cross section. Combining this with the model-independent $95\%$
confidence upper bound on the visible cross section then yields an upper bound on the charged triplet production cross section.
Note that the production of the charged triplet components (scalars) will lead to disappearing track $p_T$ and $\eta$ distributions that differ from those in chargino production (fermions). Similarly, the leading jet $p_T$ will also differ. However, the charged scalar production distributions are skewed towards higher $p_T$ and lower $\lvert{\eta}\rvert$ values, such that the acceptance times efficiency for charged triplet production is likely higher than for chargino production. Thus using the chargino acceptance times efficiency data should result in a conservative estimate for the disappearing track bound.
The triplets may also be pair produced via an intermediate
SM Higgs boson produced, increasing the total production cross section. However this production process will likely have a different jet distribution, such that the given acceptances and efficiencies likely do not apply. We will set
$\mu_\Sigma^2 = - \mSigmaN^2$ ($\lambdaSigmaH = 0$), and ignore this
production process in this section.

To interpret the interpolated disappearing track search results, we need the
lifetime and production cross section for the charged triplets. The lifetimes were calculated using
eqs.~\eqref{eq:splitting}~and~\eqref{eq:lifetime}. We utilise
\texttt{MadGraph5\_aMC@NLO}~$2.6.5$~\cite{madgraph} to evaluate the production
cross section at NLO, using an NLO compatible UFO~\cite{UFO} model file generated
using \texttt{FeynRules}~$2.3.32$~\cite{feynrules},
\texttt{FeynArts}~$3.9$~\cite{feynarts,feynrulesInterface}, and
\texttt{NLOCT}~$1.02$~\cite{NLOCT}. The charged triplet lifetime and production cross
sections are then only dependent on the mass of the triplet, and the
disappearing track searches can be used to place a lower bound on that mass.

\begin{figure}
  \centering
    \includegraphics[width=0.65\textwidth]{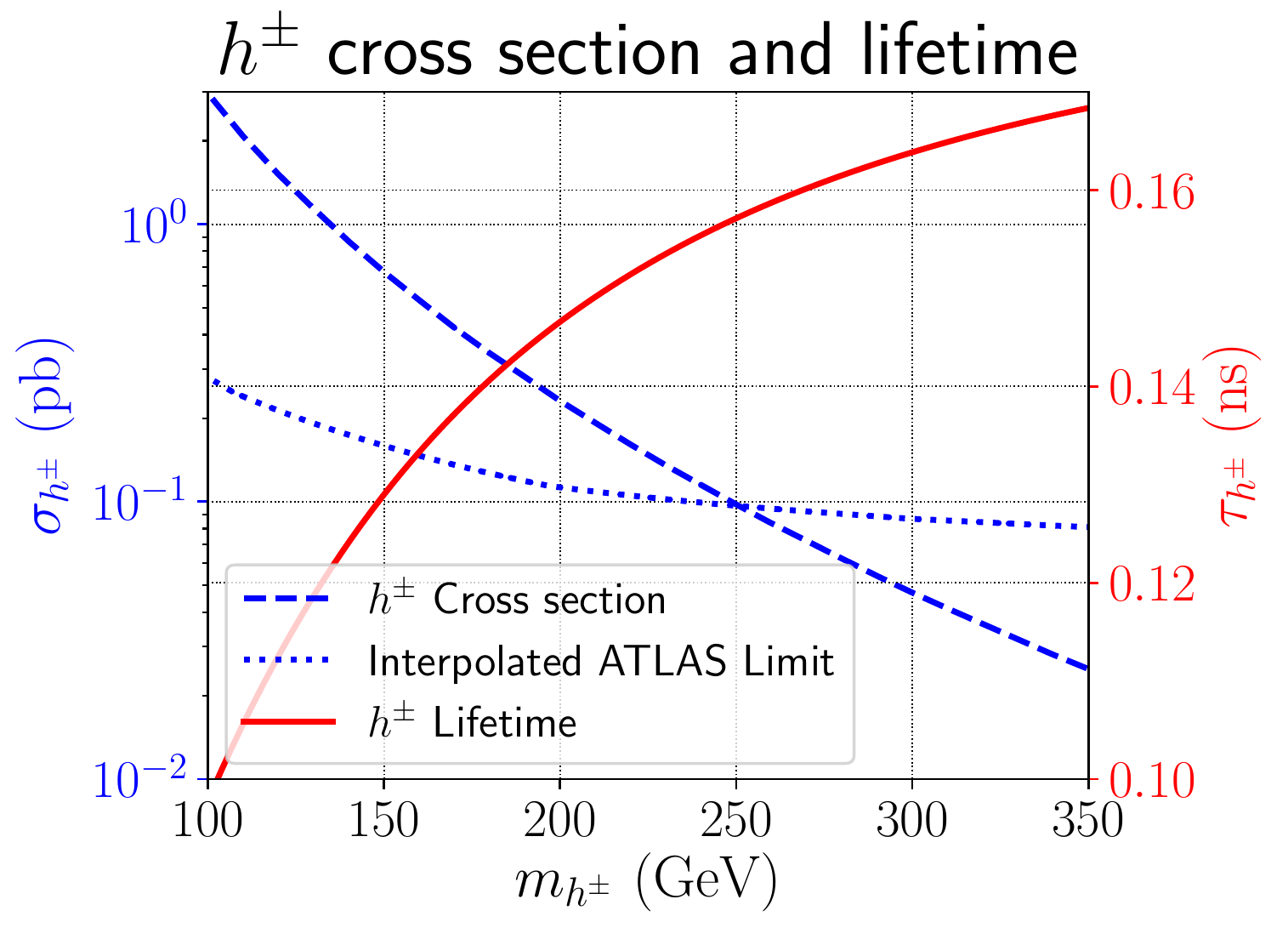}
    \caption{
  \label{fig:tripletProdAndLifetime}
  Charged triplet production cross section (dashed blue line) and lifetime
  (solid red line), along with the interpolated $95\%$ confidence upper limit on the
  chargino production cross section (dotted blue line) arising from disappearing
  tracks searches for charginos with the same mass and lifetime. The charged
  triplet production cross section intersects the upper limit at $\mSigmaC
  \approx \disappearingTrackBound\GeV$, and we take this to be the lower bound imposed by
  disappearing track searches}
\end{figure}

The resulting cross sections, interpolated limit, and lifetimes are shown in
figure~\ref{fig:tripletProdAndLifetime}. The cross section drops below the
interpolated limit for masses $\mSigmaN \gtrsim \disappearingTrackBound\GeV$, and we take this to be
the lower bound on stable triplets arising from disappearing tracks. While LEP has searched for displaced vertices in the context of SUSY searches for chargino pair-production, due to the smaller cross-section for scalar production and threshold effects, the limits from these searches for scalars are likely to be less than the 100~GeV usually stated~\cite{Pierce:2007ut,Egana-Ugrinovic:2018roi}.

This bound is very sensitive to the lifetime of the charged triplet component,
which itself depends on the mass splitting of the charged and neutral components of the triplet. The
lifetime of fermionic multiplets decaying due to radiative mass splitting has
been found to change significantly when performing a two-loop mass splitting
calculation~\cite{TwoLoopSplit1,TwoLoopSplit2}. In the fermionic case, the mass
splitting decreases and the lifetime goes up, which is favourable for the reach
of disappearing track searches. Reliably excluding the triplet would require a
precise calculation of the scalar two-loop radiative mass splitting, which is
beyond the scope of our analysis.
Additionally, note that the lifetime of the charged triplet decreases with decreasing mass. This is a result of the  fact that the one-loop radiative mass splitting, eq.~\eqref{eq:splitting}, is larger for smaller triplet masses.\footnote{This is not the case for fermionic multiplets. The fermionic mass splitting is smaller for smaller masses, such that the lifetime increases.} Thus, for some mass less than $100\GeV$ the lifetime will be too short to leave disappearing tracks, and will not be constrained by these analyses. As the available ATLAS disappearing track data only goes down to chargino masses of about $100\GeV$, it is not clear at what mass the decrease in lifetime overpowers the increasing production cross section.

\subsection{Dark Matter Direct Detection}
\label{sec:DM}

The real SU(2) triplet scalar thermal dark matter model has been studied
extensively~\cite{StrumiaMinimalDM,StrumiaSommerfeld,StrumiaCosmic,MultipletEWPTDM,TripletDM1,LHCTripDM,TripDMFootprint}. The annihilation into weak
gauge bosons requires that the triplet have a mass $\mSigmaN \sim 2 \TeV $
in order to obtain the right relic density. Inclusion of annihilation via the SM
Higgs necessitates an even larger mass. Hence a triplet with $\mSigmaN
\lesssim 500 \GeV$, as required by our constraints, will only ever constitute a
small fraction of the relic density. However, if we require $\mu_\Sigma^2>0$,
the triplet will have a large coupling to the SM Higgs. This coupling provides the dominant contribution to the
nuclear scattering cross section, and thus constrains the triplet even for very small
relic abundances.

In order to investigate this bound in more detail we utilise
\texttt{MicrOMEGAS}~$5.0.8$~\cite{micromegas} to evaluate the triplet relic abundance. We normalise the relic abundance by the dark
matter density measured by the Planck collaboration~\cite{planck2018}, $\Omega_{\mrm{DM}} h^2
= 0.12 $. The \texttt{MicrOMEGAS} results were verified by comparison with results obtained
using \texttt{MadDM}~$3.0$~\cite{maddm} and they were found to be in good agreement.
However, it is important to note that neither \texttt{MicrOMEGAS} nor \texttt{MadDM} include the Sommerfeld enhancement. The Sommerfeld enhancement arises due to the attractive potential between two DM particles resulting in an increase in the DM annihilation rate, with a corresponding decrease in the relic density. The effect is suppressed if the electroweak symmetry is broken and the weak gauge bosons gain masses comparable to the DM mass. Given that freeze-out typically occurs at temperatures $T_f \sim m_{\mathrm{DM}}/25$ and as we are interested in triplets with $\mSigmaN < 500 \GeV$, which implies $T_f \lesssim 20 \GeV$, we expect the electroweak symmetry to have been broken by the time the triplets freeze out.  However, even with massive gauge bosons, the Sommerfeld effect can still reduce the relic density by $15$--$30\%$ for triplets with masses $\mSigmaN =400$--$1000\GeV$~\cite{StrumiaSommerfeld}. We will not perform a rigorous calculation accounting for the Sommerfeld enhancement and will simply note that there is a $\sim 15\%$ uncertainty on the relic density and resulting DM detection exclusion plots.

In addition to neglecting the Sommerfeld enhancement, we also ignore bound state effects as they are negligible for the parameter-space that we consider. Furthermore, we also utilise the zero-temperature mass for the triplet during the relic density calculation. If the triplet's mass at zero temperature arises primarily through the Higgs VEV, its mass may change significantly in the early universe. However, as we expect freeze-out to occur at $T_f \lesssim 20 \GeV$, we expect $v_H$, $\mSigmaN$, and $\mSigmaN$ to be close to their zero temperature values, such that this is a minor correction. This approximation is motivated by noting that in the SM, there is a crossover transition at $T_c \sim 160 \GeV$~\cite{SMEWPT}, with the SM Higgs VEV approximately decreasing as $v_H(T)\sim v_H(0) \sqrt{1 - \frac{T^2}{T_c^2}}$. Thus at freeze-out one might reasonably expect $v_H(T_f)/v_H(0) = 0.99$, such that using the zero-temperature value for the Higgs VEV at $T\lesssim20\GeV$ is a reasonable approximation in the SM. We assume this approximation remains reasonable despite changes to the electroweak phase transition due the addition of the triplet. A precise determination of the relic density would require a proper calculation for the phase transition for each parameter point in order to obtain the correct temperature dependent masses.

\begin{figure}
  \centering
        \includegraphics[width=0.99\textwidth]{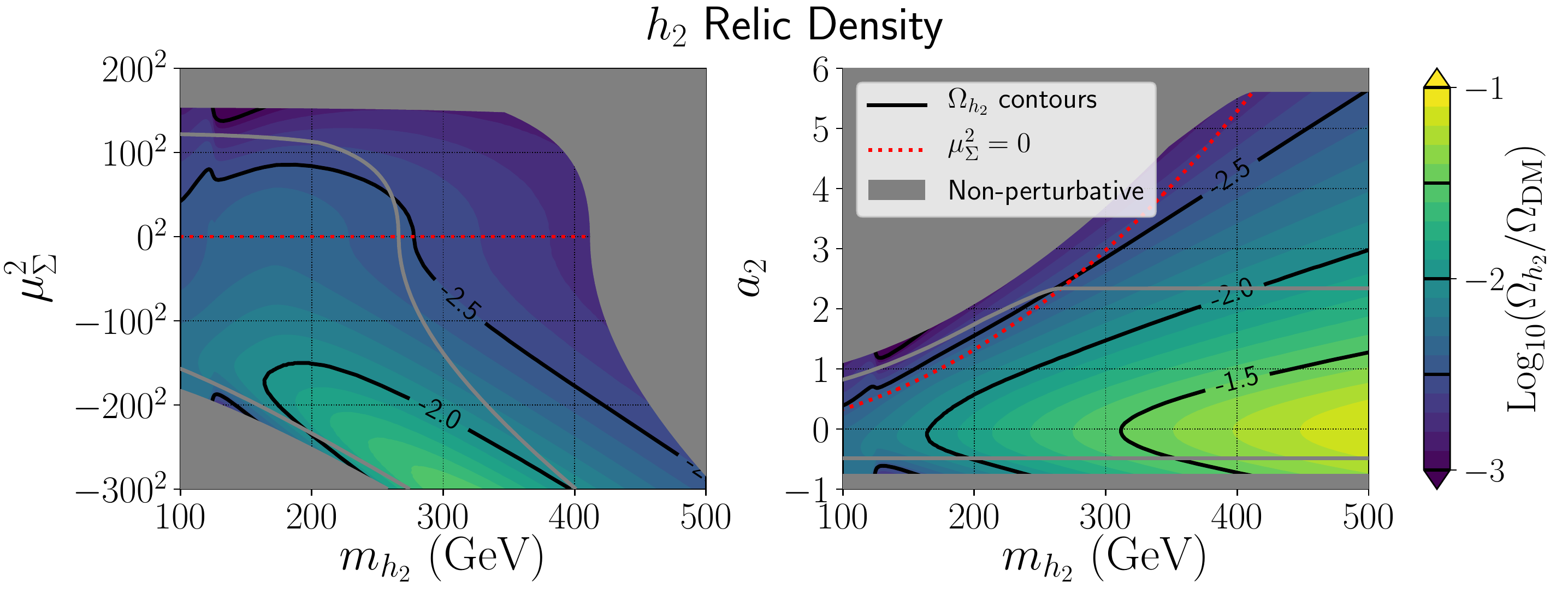}
        \caption{
      \label{fig:DM}
      The log of triplet dark matter relic density as a function of $\mSigmaN$ and either $\mu_\Sigma^2$ (left) or $\lambdaSigmaH$ (right), normalised to the observed value from Planck. The solid grey region is the parameter-space where the scalar couplings become non-perturbative at energies $\Lambda < 1 \TeV$, and the solid grey line shows where this contour would be if the cutoff energy is increased to $\Lambda=10^6\GeV$. As a function of $\lambdaSigmaH$, the relic density reaches a maximum at $\lambdaSigmaH = 0$ ($\mu_\Sigma^2 =  - \mSigmaN^2$), such that the $\SigmaN \SigmaN \rightarrow \HN \HN $ annihilation rate is zero. The red dotted line is the $\mu_\Sigma^2 = 0$ contour.}
\end{figure}

The resulting relic densities are shown in figure~\ref{fig:DM} as a function of $\mSigmaN$ and either $\mu_\Sigma^2$ or $\lambdaSigmaH$. Unless $\mSigmaN \gtrsim 500 \GeV$ and $\lambdaSigmaH \sim 0$ ($\mu_\Sigma^2 \sim  - \mSigmaN^2$), such that the annihilation rate into two SM Higgs bosons is small, the neutral triplet makes up less than $10\%$ of the total
dark matter density. The slight jump in relic density for $\mSigmaN < 125\GeV$ occurs due to the
kinematic suppression of the $\SigmaN \SigmaN \rightarrow \HN \HN$ annihilation channel, leading to a larger relic density.

\begin{figure}
  \centering
  \includegraphics[width=0.95\textwidth]{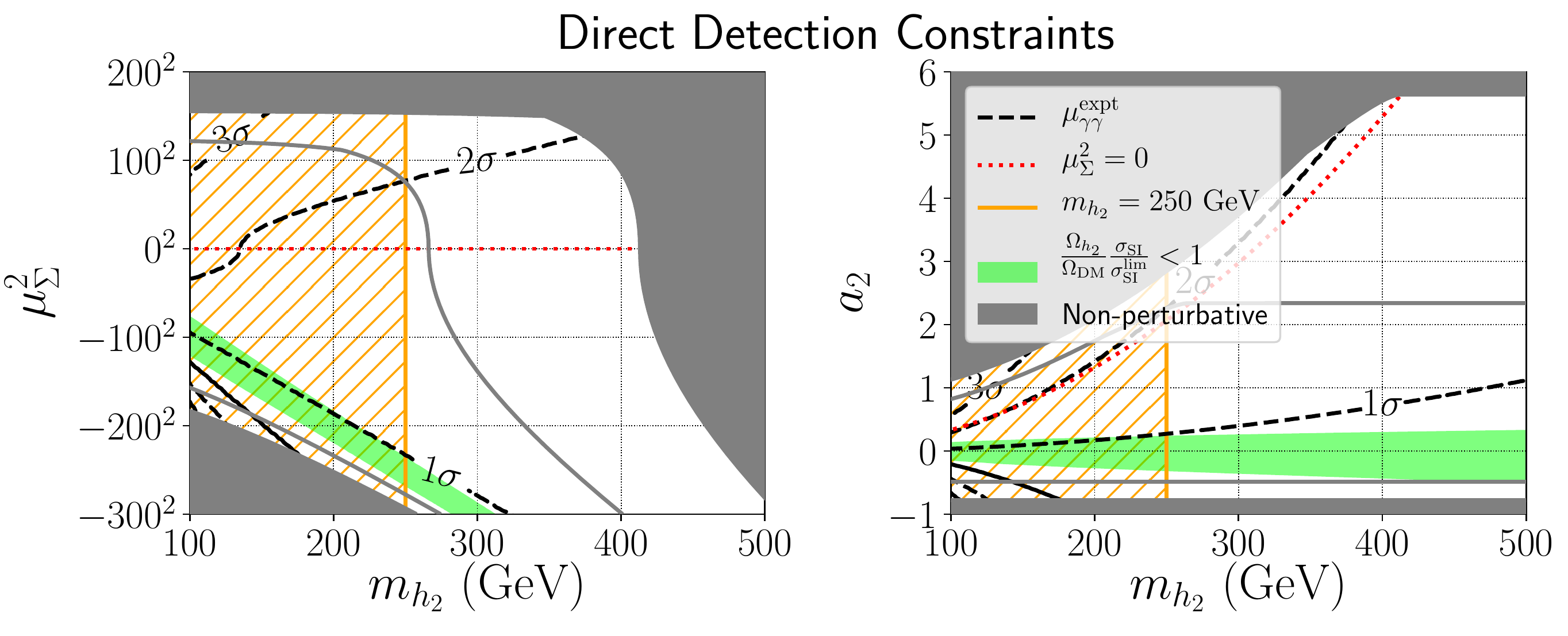}
          \caption{
      \label{fig:stableconstraints}
      The region of parameter space consistent with the current XENON1T direct detection limit as a function of $\mSigmaN$ and either $\mu_\Sigma^2$ (left) or $\lambdaSigmaH$ (right). The allowed parameter-space is shown as a green band. The solid grey region is the parameter-space where the scalar couplings become non-perturbative at energies $\Lambda < 1 \TeV$, and the solid grey line shows where this contour would be if the cutoff energy is increased to $\Lambda=10^6\GeV$.
      The black lines indicate the diphoton rate contours as in
      Fig.~\ref{fig:diphoton}.
      The hashed orange region is the lower bound on $\mSigmaN$ from disappearing
      tracks that was obtained in section~\ref{sec:dissaparingTracks}. The red dotted line is the $\mu_\Sigma^2 = 0$ contour.
      }
\end{figure}

The spin-independent (SI) nuclear scattering cross section $\sigma_\mrm{SI}$ is then obtained using the formulae given in ref.~\cite{MJRMMultiplets}, which takes into account the one-loop scattering cross section generated by $W^\pm$ box-diagrams. The cross section is then compared to the XENON1T~\cite{XENON1T1Y} $90\%$-confidence upper bound on the SI
scattering cross section $\sigma_{\mrm{SI}}^{\mrm{lim}}$, after scaling to
account for the fraction of the density of DM that is made up of $\SigmaN$.
Figure~\ref{fig:stableconstraints} shows the constraints from the XENON1T
experiment, along with the lower bound imposed by disappearing track searches. A stable triplet with $\mu_\Sigma^2>0$ is ruled out by dark
matter direct detection constraints. The only region allowed is a strip
where $\mu_\Sigma^2 \sim - \mSigmaN^2$, corresponding to $\lvert \lambdaSigmaH \rvert
\lesssim0.5$, where the triplet coupling to the SM Higgs is small. This is shown
as a green band in figure~\ref{fig:stableconstraints}. As the rate for DM
self-annihilation is proportional to the number density squared, the
annihilation rate is very low for these relic densities. Hence, there are no
constraints from dark matter indirect detection experiments.
Inclusion of the Sommerfeld enhancement would result in a slightly larger allowed region.

Higher representation $SU(2)$ multiplets are also strongly constrained by dark
matter direct detection constraints, forcing the coupling between the SM Higgs
and scalar electroweak multiplet dark matter to be
small~\cite{MorriseyTwoStep,MJRMMultiplets}. These direct detection constraints
are not always applied even when they rule out a significant region of the
benchmark points considered, as is the case for
refs.~\cite{StepInto,TwoStep,tripletLattice}. These models then require either
allowing for $\mu_\Sigma^2<0$ and tuning the DM-Higgs coupling to be small, which
is unfavourable for two step phase transition models, or breaking the
$\mathbb{Z}_2$ which stabilises the DM, so that it is not a DM candidate any
more.\footnote{If the scalars are long lived but unstable, the disappearing track
  constraints will still apply.} Breaking the symmetry is
straightforward in the case of the scalar triplet. No additional particle
content is needed as allowing a non-zero $\aSigmaH$ coupling breaks the
symmetry. This $\mathbb{Z}_2$ breaking coupling can be very small, such that it
will not significantly change the results of phase transition studies. However,
one must then contend with new constraints arising from other collider searches,
and it is to this possibility that we now turn our attention.

\section{Unstable Triplet Phenomenology}
\label{sec:unstablePheno}

\subsection{Production Processes}
\label{sec:production}
\begin{figure}
  \centering
  \begin{subfigure}[b]{0.19\textwidth} 
  \includegraphics{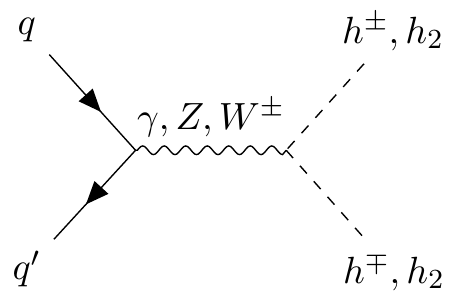}
\caption{}
\label{fig:FeynDY}
\end{subfigure} \qquad \qquad
\begin{subfigure}[b]{0.19\textwidth} 
  \includegraphics{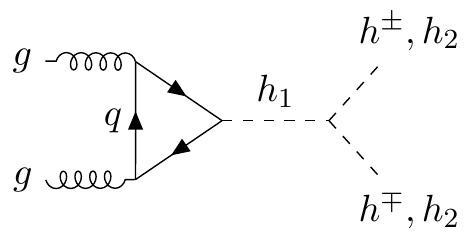}
\caption{}
\label{fig:FeynggF}
\end{subfigure} \qquad  \qquad  \qquad
\begin{subfigure}[b]{0.19\textwidth} 
  \includegraphics{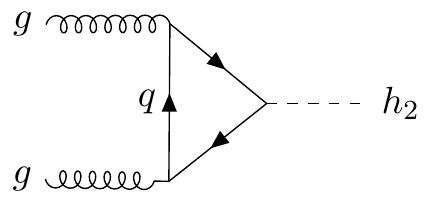}
\caption{}
\label{fig:FeynggFSingle}
\end{subfigure}
  \caption{
    \label{fig:prodFeynman}
    Feynman diagrams showing the primary production processes for the new
    scalars, including Drell-Yan
    pair production (\subref{fig:FeynDY}), pair production via an intermediate
    off-shell $h_1$ (\subref{fig:FeynggF}), and single
    $h_2$ production via $gg$F (\subref{fig:FeynggFSingle}).}
\end{figure}

The primary production processes for the SM Higgs boson at the LHC are via gluon-gluon
fusion ($gg$F) and vector-boson fusion (VBF). However, neither of these processes
will lead to appreciable $h_2$ production. This is due to the fact that the
coupling to the heavy quarks involved in $gg$F is suppressed by a factor of $\sin
\theta_N$, leading to a significantly smaller production rate. Additionally, the
$W W h_2$ and $Z Z h_2$ vertices necessary for VBF arise due to neutral scalar
mixing (suppressed by $\sin \theta_N$) or via the triplet's VEV (suppressed by
$v_\Sigma/v_H$). Other SM-Higgs production mechanisms are similarly suppressed.
Hence, unless $h_1$ and $h_2$ are nearly degenerate, such that there is a sizeable
mixing angle, single $h_2$ production will be several orders of magnitude
smaller than SM Higgs production cross sections. Single $h^\pm$ production will
similarly be suppressed by factors of $\sin \theta_C$ and $v_\Sigma/v_H$. As a
result, the primary production mechanism for the new scalars is via neutral or
charged current Drell-Yan pair production. Additionally, pair production via an
intermediate off-shell SM Higgs may contribute significantly. In the SM, Higgs
pair production is suppressed due to the small cubic coupling $\lambda_H v_H$,
and due to the interference of the box and triangle diagrams~\cite{Dolan:2012rv}. However, in our
scenario the coupling $\lambdaSigmaH v_H$ may be large and the
interfering box diagram is suppressed by a factor of $\sin^2 \theta_N$. Thus
production via an off shell $h_1$ can form a significant contribution for large
$\lambdaSigmaH$. We will therefore include pair production via an off-shell
intermediate $h_1$ produced through $gg$F. All $h_1$-style production processes
will contribute in such a manner. However, as $gg$F is the dominant production
process for single $h_1$ and as Drell-Yan pair production dominates anyway,
neglecting other off-shell $h_1$ pair production diagrams will have no
significant effect on the results. Feynman diagrams for the dominant production
processes are shown in figure~\ref{fig:prodFeynman}.

\begin{figure}
  \centering
    \includegraphics[width=0.65\textwidth]{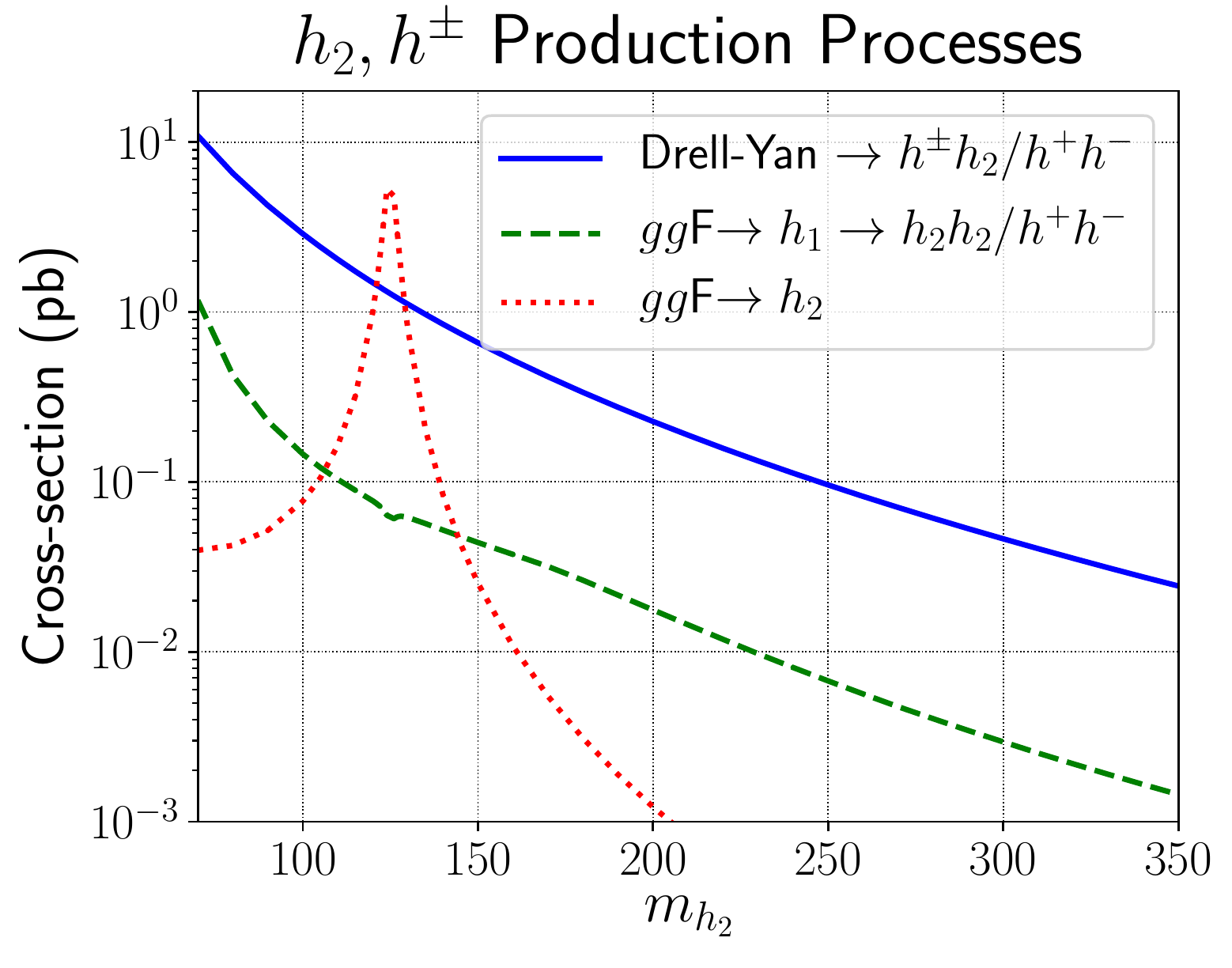}
    \caption{
  \label{fig:tripletProd}
  The total $h_2, h^\pm$ pair
  production cross sections via Drell-Yan processes (solid blue line) or via an intermediate off-shell $h_1$ (green dashed line), as well as the single $h_2$ production cross
  section via $gg$F (dotted red line). The cross sections are shown for $\mu_\Sigma^2 = 200^2 \ \mathrm{GeV}^2$ and $v_\Sigma=1$.
      The peak in the single $h_2$ production cross at $m_{h_2} \approx
      125\ \GeV$ occurs due to an increase in the neutral scalar mixing angle when the two neutral scalars are nearly degenerate, and not due to an $s$-channel resonance.
  }
\end{figure}

Figure~\ref{fig:tripletProd} shows the pair production cross sections for the
new scalars via Drell-Yan or via an intermediate off-shell $h_1$, in addition to
single $h_2$ production via $gg$F. The cross sections were obtained using
\texttt{MadGraph5}. The Drell-Yan cross section was evaluated with NLO
QCD corrections, while the off-shell $h_1$ and $gg$F $h_2$ production cross
sections are loop induced processes evaluated at leading order. As argued
earlier, the cross section for the production of a single $h_2$ is via $gg$F is
suppressed by $\sin \theta_N$, such that it is large only when $m_{h_1}\approx
m_{h_2}$. Pair production dominates away from this region, and will always dominate if $v_\Sigma\lesssim 0.5 \GeV$. Furthermore, pair
produced $h_2$ lead to multi-gauge boson events with significantly smaller
backgrounds, and as a result we focus on pair production at colliders.

Note that $gg$F Higgs production increases significantly with the inclusion of
higher order corrections, with a $k$-factor of around 3 at
$\rm{N}^3\rm{LO}$~\cite{Anastasiou:2015ema}. The cross sections shown in
figure~\ref{fig:tripletProd} are unmodified. Even with the correction, single
$h_2$ production remains subdominant for most of the parameter-space. However, this
raises a concern that the higher order corrections to the pair production
through intermediate off-shell $h_1$ are similarly significant. Naively, as the
QCD component of the $gg$F single Higgs and new scalar $gg$F pair production are
the same, one might expect a $k$-factor of $k\sim 3$. In contrast, the
$k$-factor for SM Higgs pair production is $k\sim
2$\cite{Dawson:1998py,deFlorian:2016uhr,Grazzini:2018bsd}. As mentioned before, this
process is different as it features an additional interfering box diagram. It is
unclear which $k$-factor is more readily applicable to the new scalar $gg$F pair
production process. We take the lower of the two and scale this cross section by
a $k$-factor of $k=2$. With the $k$-factor correction, pair production via an
intermediate $h_1$ results in a $10$--$20$\% increase in the overall pair
production cross section. Furthermore, it is the only source of $h_2 h_2$ pairs,
as they are not produced via neutral-current Drell-Yan processes.

\subsection{Decay Channels}
\label{sec:unstabledecays}

The $h_2$ and $h^\pm$ scalars have three means of decaying:
\begin{itemize}
\item Decay via mixing with the SM Higgs or charged Goldstone into fermions and
  gauge bosons. For the $h_2$, these partial widths are suppressed relative to
  SM Higgs decays by a factor of $\sin \theta_{N}$, while the partial widths for
  the $h^\pm$ will be proportional to $\sin \theta_C$.
\item Decay via $v_\Sigma$ into weak gauge bosons ($W^\pm W^\mp$, $Z Z$, $W^\pm
  Z$). These partial widths are suppressed by $v_\Sigma/v_H$ relative to similar
  SM Higgs decays.
\item Decay into $h_1 h_1$ or $W^\pm h_1$. These partial widths are
  proportional to  $v_\Sigma + v_H \sin\theta_N$ and $\sin \theta_C + 2\sin
  \theta_N$, respectively.
\end{itemize}
Thus, aside from the $h^{\pm}
\rightarrow h_1 W^\pm$ channel which depends on $\theta_N$, the partial widths of the charged scalar are completely determined by its
mass (kinematics) and
$v_\Sigma$ (which fixes $\theta_C$).
The scenario for the neutral scalar is more complicated, as
$\theta_N$ is a function of $v_\Sigma$, $\mu_\Sigma^2$ and $m_{h_2}$. Additionally, one must include the $\lambdaSigma$ dependent diphoton rate.
For the purposes of the decay phenomenology, changing $\mu_\Sigma^2$
affects the size of the neutral scalar mixing angle $\theta_{N}$.
In particular, note that from eq.~\eqref{eq:vevs}, if $\mu_\Sigma^2 = v_\Sigma^2
\lambdaSigma$, then $v_\Sigma = \frac{\aSigmaH}{2 \lambdaSigmaH}$ such
that the off-diagonal term in the scalar mass mixing matrix disappears and we
get $\theta_N = 0$. Hence when $\mu_\Sigma^2$ is small, of the order of a few
$\mathrm{GeV}^2$, the neutral scalar mixing angle $\theta_{N}$ will also be very
small and the decays of the neutral triplet will be dominated by decays into weak gauge
bosons $h_2 \rightarrow W^\pm W^{\mp (*)}$.
Conversely, a larger $\lvert  \mu_\Sigma^2 \rvert$ corresponds to a larger
$\theta_N$. Finally, as both $\theta_N$ and $\theta_C$ are both proportional to
$v_\Sigma$, the triplet VEV sets the overall size of the widths and has very little
impact on the branching fractions.  

To obtain the partial widths for decays into fermions and gluons arising from
mixing with the SM Higgs, we utilise the \texttt{HDECAY}~$6.511$~\cite{hdecay}
package. The $h_2$ partial widths for decays into fermions and gluons are those
of a SM Higgs of mass $m_{h_2}$ scaled by $\sin^2 \theta_{N}$. For the diphoton
rate $\Gamma^{\mathrm{\Sigma SM}}_{h_2 \rightarrow \gamma \gamma}$ we use the
analytic formulae given in eq.~\eqref{eq:TripDiphoton}. We do not include the $h_2
\rightarrow Z \gamma$ decay. The partial widths for the decay of $h^\pm$ into
fermions are obtained from the partial widths of a charged Higgs in a type-I
2HDM with $\tan{\beta} = \frac{1}{\tan{\theta_C}}$ and $\sin(\alpha)=0$, as given
by \texttt{HDECAY}. The other decays into scalars and electroweak gauge bosons
were obtained automatically by \texttt{MadWidth}~\cite{madwidth}, a component of \texttt{MadGraph5}.

\begin{figure}
  \centering
    \includegraphics[width=0.98\textwidth]{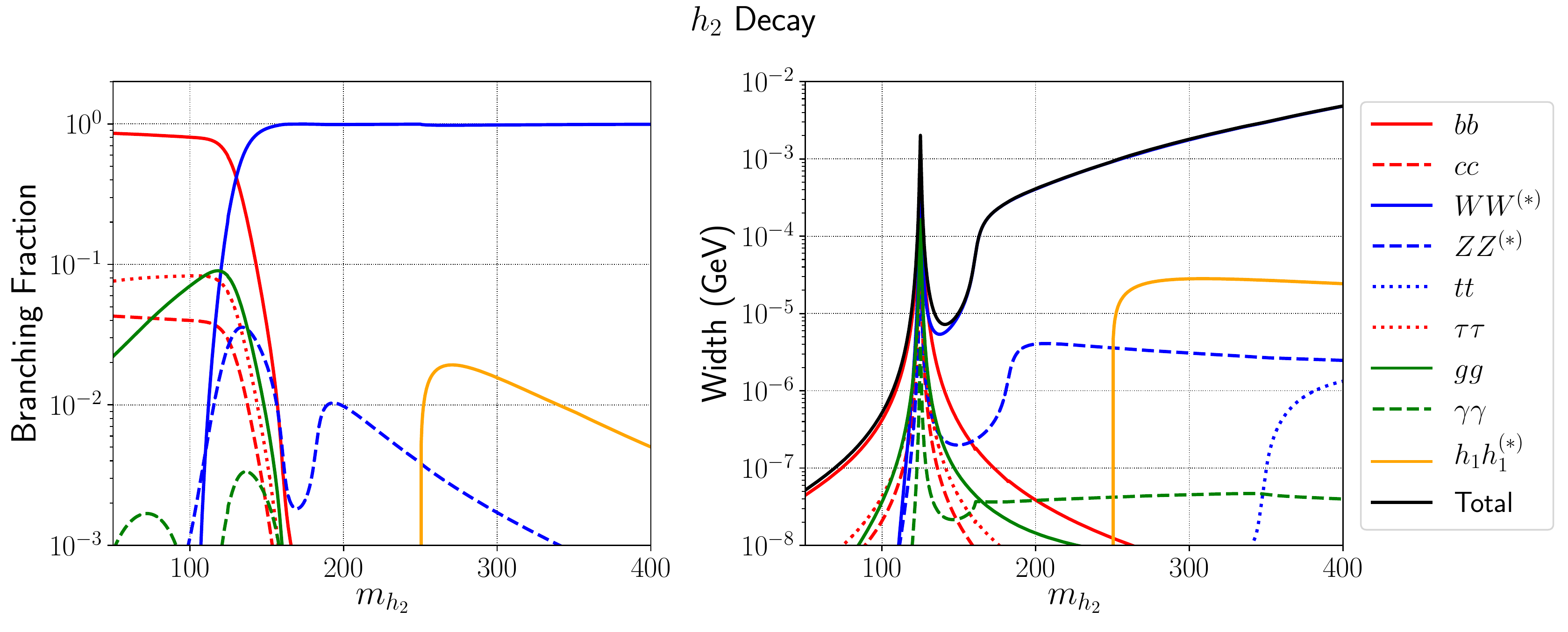}
    \caption{
      \label{fig:h2Decay}
      Branching fractions (left panel) and partial widths  (right panel) for the decay of $h_2$, for
      $\mu_\Sigma^2= 100^2\ \mathrm{GeV}^2$, $v_\Sigma=1\ \GeV$, and $\lambdaSigma=1$. 
      The peak at $m_{h_2} \approx
      125\ \GeV$ occurs due to an increase in the neutral scalar mixing angle when
      the two neutral scalars are nearly degenerate.
      }
\end{figure}

\begin{figure}
  \centering
    \includegraphics[width=0.98\textwidth]{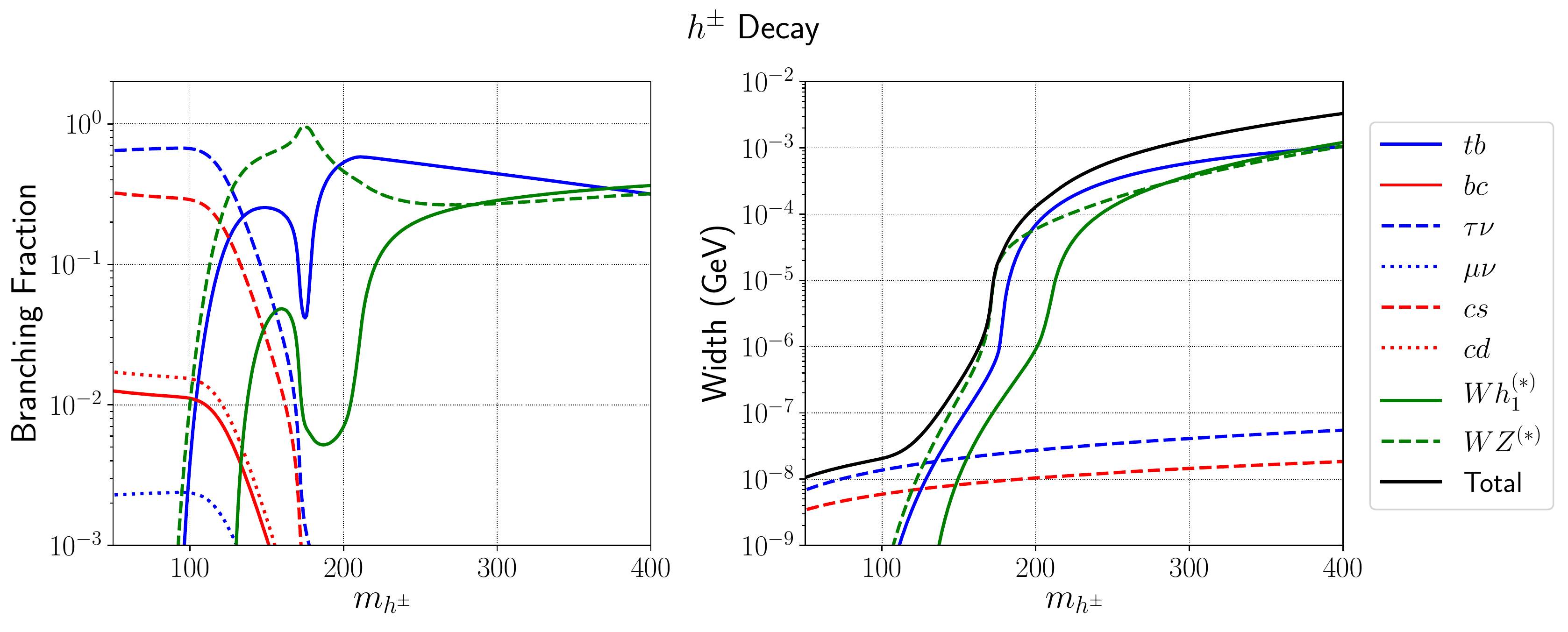}
    \caption{
      \label{fig:hcDecay}
      Branching fractions (left panel) and partial widths (right panel) for the decay of $h^+$ for
      $\mu_\Sigma^2= 100^2\ \mathrm{GeV}^2$, $v_\Sigma=1\ \GeV$, and $\lambdaSigma=1$. }
  \end{figure}

The resulting branching fractions and partial widths are shown in
figures~\ref{fig:h2Decay}~and~\ref{fig:hcDecay} for the triplet-like neutral and
charged scalar, respectively. The $h_2$ width features a resonance at
$m_{h_2}\approx 125\GeV$ due to a large neutral scalar mixing angle
$\theta_N$. However, the branching fractions are relatively smooth and instead
feature a transition between fermionic and electroweak decays due to kinematic
suppression. For small masses, the $h_2$ decay primarily into
$b\bar{b}$, $\tau^+ \tau^-$, and $c \bar{c}$. Conversely, for larger masses they decay
primarily into $W^\pm W^\mp$ and $h_1 h_1$, with the branching ratio into the
latter being strongly dependent on $\mu_\Sigma^2$. The charged scalar will decay
mostly into $\tau \nu$ or $c s$ fermions if $m_{h^\pm}\lesssim 120 \GeV$, and
into $t b$, $W^\pm h_1$, or $W^\pm Z$ pairs when heavier.
Note that if we had fixed $\lambdaSigmaH$ instead of setting $\mu_\Sigma^2 = 100^2\  \mathrm{GeV}^2$ the behaviour of the branching fraction would be significantly different. In particular, if $\lambdaSigmaH$ is negative, then $\mathrm{Br}(h_2 \rightarrow Z Z)$ can become large.
Additionally, if $
\mu_\Sigma^2 \sim v_\Sigma^2 \lambdaSigma$ such that $\theta_N$ is small and the SM Higgs mixing-induced decays into two fermions are suppressed, and $m_{h_2} \lesssim m_W$, so that the decays
into weak gauge bosons are kinematically suppressed, then the $h_2$ diphoton
branching fraction can become significant.

Note that our choice to fix the cubic term $\aSigmaH$ as a function of
$v_\Sigma$ significantly affects the behaviour of the widths as a function of
mass. If we had instead selected a value for $\aSigmaH$ and used that to
fix $v_\Sigma$ (leading to $v_\Sigma \propto 1/m_{h_2}^2$), the
partial widths would decrease as the mass of the triplet becomes very large.
However, as mentioned earlier, varying $v_\Sigma$ scales the overall widths
without affecting the branching ratios. Hence, the phenomenological results
would be the same.

\subsection{Collider Searches}

There are a range of ATLAS and CMS analyses searching for exotic scalars. However, these searches generally focus on single scalar production. Examples include searches for new neutral scalars decaying into $\gamma \gamma$~\cite{diphotonLowATLAS,diphotonLowCMS,diphotonHighATLAS,diphotonHighCMS} or $\tau^+ \tau^-$~\cite{h2-tau-tau-CMS}, and searches for new charged scalars decaying into $\tau \nu_\tau$~\cite{hc-tau-nu-ATLAS,hc-tau-nu-CMS}, $t b$~\cite{hc-tb-CMS,hc-tb-ATLAS}, or $W Z$~\cite{hc-WZ-ATLAS,hc-WZ-CMS}.
In our scenario the dominant source of new scalars is via pair production. While pair production may lead to signal events in these searches, dedicated pair production searches would have significantly lower backgrounds, and thus, would be significantly more constraining. Taking the constraints on the production cross section times branching fractions obtained by these analyses and directly interpreting them as constraints on the pair-production cross-section times branching fractions, we find that none of these searches constrain the $\Sigma$SM.\footnote{Except for a small region of parameter-space with a large diphoton branching fraction, which is discussed in more detail in the next section.} Note that this interpretation neglects the details of the analyses, i.e., ref.~\cite{hc-tau-nu-ATLAS} specifically searches for, and places constraints on, $h^\pm b \bar{b} W^\mp$ production, not general $h^\pm$ production. However, as the bound on the cross sections is too weak to constrain the $\Sigma$SM, a more detailed examination is unnecessary.

There are dedicated pair production searches for neutral scalars with a focus on new contributions to SM Higgs pair production~\cite{CMS-HH,ATLAS-HH}. The signal regions in these analyses will constrain our model. However, as mentioned previously, the only source of neutral $h_2 h_2$ pairs is via an off-shell $h_1$. While this process can give a $\sim 10\%$ correction to the overall pair production cross section for large $\mu_\Sigma^2$, the cross section is too small to be constrained by these searches. Additionally, if the new charged scalars are heavy, $h_1 h_1$ pairs could be produced via $h^+ h^- \rightarrow h_1 h_1 W^+ W^-$. However, once again the cross section and branching fractions are too small to be constrained by current SM Higgs pair production searches. 

As pointed out by refs.~\cite{ColliderPairProd,TripletGravWaves}, there is a lack of dedicated searches for pair production involving charged scalars at $13\TeV$. In particular, there are no recent searches with $t \bar{b}, \bar{t} b$ final states, which might arise in the triplet model via $h^+ h^-$ pair production if $m_{h^\pm}>m_t$. Similarly, there are no recent searches with $t \bar{t}, t \bar{b} (\bar{t} b)$ or $b \bar{b}, t\bar{b} (\bar{t} b)$ final states, which may arise in $h_2 h^\pm$ pair production. The latter of these final states is explored in ref.~\cite{TripletGravWaves}. However, for $m_{h_2}\gtrsim 150\GeV$ our branching fraction $\mathrm{Br}\left(h_2 \rightarrow b \bar{b}\right)$ becomes too small for this final state to constrain the minimal triplet model.

There are other LHC searches that feature similar final states that can be used to constrain the $\Sigma$SM. In particular, note that when the triplets are light ($m_{h_2} \lesssim 110\GeV$) $h_2 h^\pm$ production can result in $ \tau \tau \tau \nu_\tau$ pairs. On the other hand, for heavy triplets, processes such as $h^+ h^- \rightarrow W^+ W^- Z Z$ or $ h_2 h^\pm \rightarrow W^+ W^- W^\pm Z$ can lead to a large number of leptons if some of the weak gauge bosons decay leptonically. Therefore searches featuring multilepton signal regions can be used to place constraints on the $\Sigma SM$.

\subsection{Collider Constraints}

We utilise the \texttt{CheckMATE}~$2.0.26$~\cite{checkmate} package in order to examine the constraints arising from multilepton collider searches. \texttt{CheckMATE} compares simulated
collider events against a range of CMS and ATLAS analyses and determines
whether a given model is excluded. We utilise
\texttt{MadGraph5} to generate parton level pair production events, with the
production processes described
in section~\ref{sec:production}. These events are showered by
\texttt{Pythia}~$8.230$~\cite{pythia3} and are then run
through the \texttt{Delphes}~$3.4.1$~\cite{delphes} detector simulation using
the \texttt{CheckMATE} interface. \texttt{CheckMATE} then evaluates the
$\mrm{CL}_{\mrm{S}}$~\cite{CLS} value for every signal region in each of the
implemented CMS and ATLAS analyses and uses the most sensitive signal region to
determine whether a model is excluded or not. These tools are dependent on a variety of other packages and tools~\cite{hepmc,fastjet2,antikt,madfks,madloop,fastjet1,scalePDFuncertainties,lhapdf6,MG5LoopInduced}.

The most constraining analyses are generally ATLAS or CMS searches for charginos
and neutralinos with multilepton final states, specifically the searches in
refs.~\cite{CMSSUS16039}~and~\cite{atlas170807875}, each using
$36\ \rm{fb}^{-1}$ of data taken at 13~TeV. Additionally, as mentioned in previous
sections, the diphoton branching fraction for the new scalar can be large if
$\mu_\Sigma^2 = v_\Sigma^2 \lambdaSigma$ and $m_{h_2}\lesssim m_W$. Hence these parameter points are excluded by analyses with photonic signatures, such as
ref.~\cite{atlas180203158}. Note that this region of parameter space is also excluded by direct diphoton resonance searches, which are not yet implemented in checkmate~\cite{diphotonLowATLAS,diphotonLowCMS}.

We varied the mass of the triplet-like neutral scalar $m_{h_2}$ from $70$ to
$350\GeV$ in steps of $10\GeV$. We let $\mu_\Sigma^2$ range from $-100^2$ to $200^2\ \mathrm{\GeV}^2$. For $m_{h_2} < 150 \GeV$ and $m_{h_2}\geq 150\GeV$, we let $\sqrt{\lvert \mu_\Sigma^2 \rvert}$ vary in steps of $25$ and $100\GeV$, respectively. The triplet quartic coupling and VEV were set to $1$ and
$1\GeV$, respectively. Note that setting $\lambdaSigma=1$ violates eq.~\eqref{eq:stability} for large values of $\mu_\Sigma^2$. However, $\lambdaSigma$ has negligible impact on collider phenomenology when $\mu_\Sigma^2$ is large, such that the results are independent of the choice of $\lambdaSigma$. In order to increase the fraction of generated events resulting in signal events for parameter points with $\mSigmaN \leq 100\GeV$ and $\lvert \mu_\Sigma^2 \rvert \geq 50\GeV^2$, the triplet-like scalars were forced to decay into $\tau$ leptons using \texttt{MadSpin}~\cite{madspin}. Outside of this region of parameter space all decays were allowed. Five million pair production events were generated for most parameter sets. Ten million events were generated for points near the 95\% exclusion boundary.

\begin{figure}
  \centering
    \includegraphics[width=0.74\textwidth]{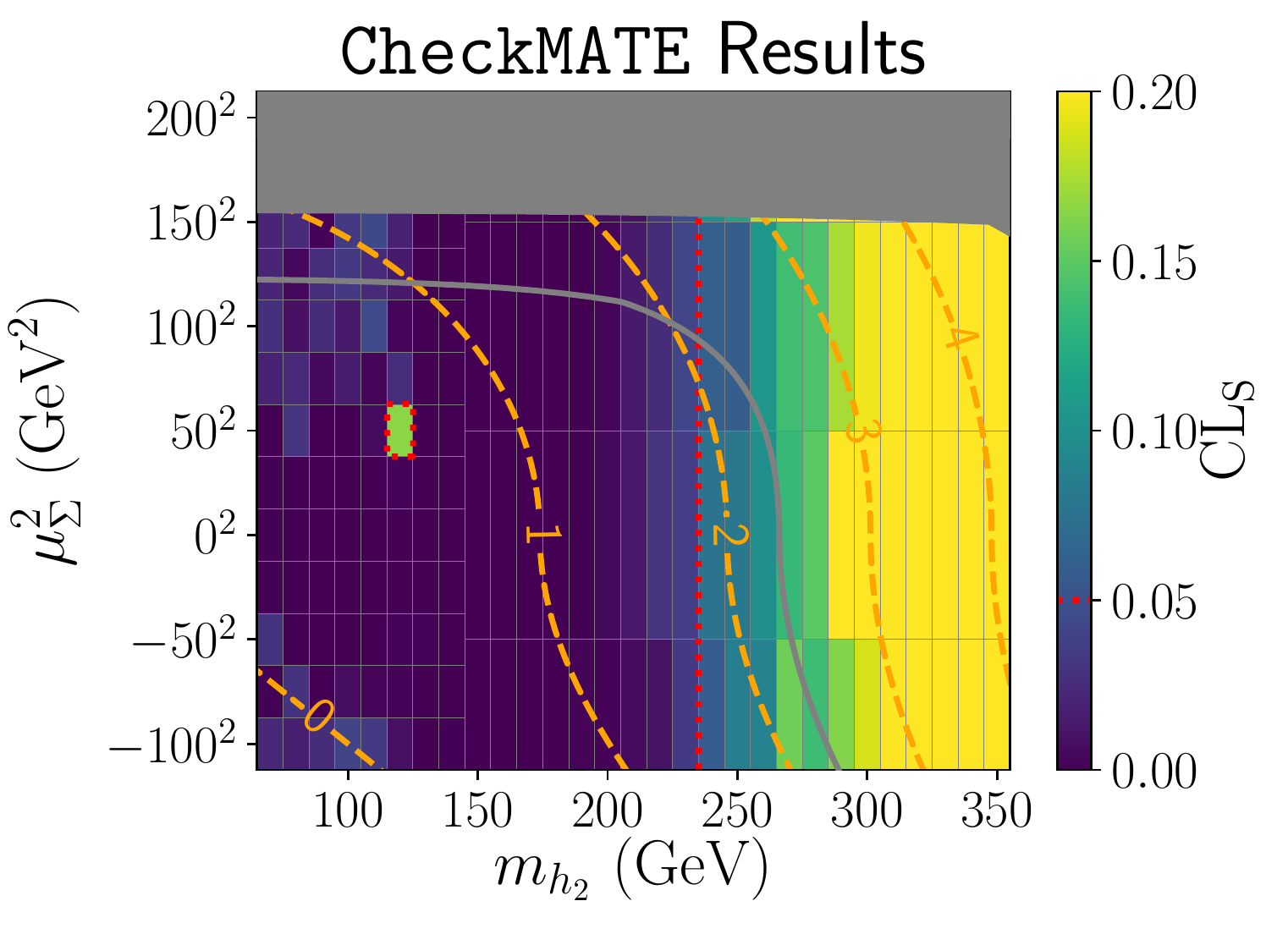}
    \caption{
      \label{fig:checkMATE}
      \texttt{CheckMATE} $\mrm{CL}_{\mrm{S}}$ exclusion values evaluated on a grid of
      masses and quadratic terms with $v_\Sigma=1\GeV$ and $\lambdaSigma = 1$. Points with $\mrm{CL_{\mrm{S}}}\leq \rlim$ are excluded at $\rconf \%$ confidence. The $\mrm{CL_{\mrm{S}}}=\rlim$
      contour is indicated by a dashed red line. 
      The dotted orange lines are contours of constant triplet-Higgs coupling. The solid grey region is the parameter-space where the scalar couplings become non-perturbative at energies $\Lambda < 1 \TeV$, and the solid grey line shows where this contour would be if the cutoff energy is increased to $\Lambda=10^6\GeV$. The perturbativity contour does not use $b_4=1$, and instead selects $b_4$ as described for figure~\ref{fig:rge}.
      }
\end{figure}

The resulting $\mrm{CL}_{\mrm{S}}$-values obtained by
\texttt{CheckMATE} are shown in figure~\ref{fig:checkMATE}. From the figure, we see that an unstable triplet-like scalar is required to have a
mass $m_{h_{2}} \gtrsim \massBound \GeV$. The one exception is a region of the parameter space near $m_{h_2} = 120\GeV$ and $\mu_\Sigma^2 =50^2\ \mrm{GeV}^2$, which is only excluded at $\problemCLs\%$ confidence. As seen in figures~\ref{fig:h2Decay}~and~\ref{fig:hcDecay}, masses near $120\GeV$ correspond to the transition between weak gauge boson and fermion pair decays, with the branching fraction of $h_2 \rightarrow \tau^+ \tau^-$ and $h^+ \rightarrow \tau^+ \nu_\tau$ decreasing. Furthermore, the rate of $h_2 \rightarrow W^\pm (W^\mp)^*$ is proportional to $4 \frac{v_\Sigma}{v_H} \cos{\theta_N} + \sin{\theta_N}$, which goes to zero near $\mu_\Sigma^2 =50^2\ \mrm{GeV}^2$ and $m_{h_2}=120\GeV$. Both of these factors combined lead to slightly fewer signal events near $\mSigmaN = 120\GeV$ and $\mu_\Sigma^2 =50^2\ \mrm{GeV}^2$. Thus this region of parameter space is not quite excluded by \texttt{CheckMATE}. Note that \texttt{CheckMATE} determines its $\mrm{CL}_{\mrm{S}}$ values using only the signal region that has the best sensitivity assuming the observed number of events match the SM prediction. This is done in order to avoid falsely excluding a model due to a downward fluctuation in the observed number of events. However, it should be noted that while the most sensitive signal region, region I04 in ref.~\cite{CMSSUS16039}, does not exclude this point, three other signal regions from the same analysis (C18, G03, G05) each individually exclude this point at 94\% confidence. We have also utilised \texttt{HiggsBounds}~$5.3.2$\texttt{beta}~\cite{HiggsBounds1,HiggsBounds2,HiggsBounds3,HiggsBounds4,HiggsBounds5} and \texttt{HiggsSignals}~$2.2.3$\texttt{beta}~\cite{HiggsSignals1,HiggsSignals2,HiggsSignals3} in order to verify that this parameter point is not separately excluded by dedicated new scalar searches or corrections to SM Higgs signals.

Recently new searches with multilepton signals have been released that utilise
up to $139\ \rm{fb}^{-1}$ of data~\cite{CMS-PAS-SUS-19-008,CMSVectorLikeLeptons,ATLAS-CONF-2019-020,Aad:2019ftg}. These analyses have not yet
been implemented in \texttt{CheckMATE}. Based on a simple scaling approximation
using the Collider-Reach tool~\cite{colliderReach}, we expect that this will
increase the lower bound on the triplet mass to above $m_{h_2} \sim \colliderReachBound \GeV$, and we expect the small allowed region to become excluded.

These collider constraints significantly restrict the parameter-space available
for novel multi-step electroweak baryogenesis models. In particular, the
parameter-space considered in ref.~\cite{StepInto}, and a significant chunk of
the parameter-space considered in ref.~\cite{tripletLattice,TripletGravWaves}, are excluded by
these constraints. We expect that other models featuring SU(2) triplet scalars decaying
in such a manner would be similarly constrained.

Note that we have only considered values of the triplet VEV that result in
short-lived triplets. In the limit where the triplet VEV approaches zero
($v_\Sigma\lesssim 10^{-4}\GeV$~\cite{MJRMTripletPheno}), the decays of the
scalars will once again resemble those in the $\mathbb{Z}_2$ symmetric case; the
$h_2$ will be stable on detector timescales and the $h^\pm$ will decay into $h_2
\pi^\pm$ or $h_2 \ell \nu$. At this point disappearing tracks will once again
constrain the triplet, though dark matter direct detection constraints are
avoided. For some small range of $v_{\Sigma}$ (or equivalently $a_1$) the decays of $h_2$ will be displaced from the primary vertex but still inside the detector. In this case both the disappearing charged track and multilepton searches will lose their efficacy. The detailed phenomenology of this intermediate regime is worth a study in its own right, and could be constrained through searches for displaced jets and leptons such as~\cite{Sirunyan:2018vlw,Aad:2019tcc}.
 Displaced vertex searches for scalars have been considered in the context of type-II seesaw models~\cite{LeftRightSeesaw}. However, we are unaware of any such search for the minimal hypercharge-zero triplet scalar model.

\section{Conclusion}
\label{sec:conclusion}

Taking to heart the notion that electroweak baryogenesis is attractive for its
testability at the LHC and prospective future colliders, we have examined the
phenomenology of light SU(2) real triplet scalars motivated by multi-step electroweak
phase transitions. We have demonstrated that such scalars are nearly excluded if
they are stable. The only region of parameter-space still allowed is where the magnitude of the Higgs portal coupling $< 1$,
which is unsuitable for a two-step
electroweak phase transition. This constraint can be avoided by breaking the
$\mathbb{Z}_2$ symmetry that stabilises the neutral component of the triplet,
allowing it to decay. However, depending on the lifetime of the charged triplets, one
must then contend with either disappearing track or multilepton searches at
colliders. These searches constrain the mass of the triplet to be at least
$\disappearingTrackBound$ or $\massBound\GeV$, respectively. It may be possible that there is a region of
parameter-space in-between the two extremes where both search types
lose sensitivity. However, this likely requires a finely tuned selection for the
triplet VEV. 

It should be noted that electroweak baryogenesis in the presence of a real triplet scalar extension of the Standard Model requires
particle content beyond the $\Sigma$SM. In particular, the $\Sigma$SM provides
no additional sources of CP violation. Therefore, any realistic
electroweak baryogenesis model will necessarily feature additional particles
which may couple to the real scalar triplet, as is the case in ref.~\cite{TwoStep}.
In addition to the phenomenology introduced by the new particle content, the
decay channels of the triplets would likely also be modified such that the
results obtained here will not be directly applicable. However, the collider
constraints on the triplet parameter-space will likely be similarly restrictive. Alternatively, the constraints imposed on the scalar potential could be relaxed
by considering further extensions of the scalar sector. This might allow for a large negative $\mu_\Sigma^2$ term, such that the triplet can be heavier at zero temperature. 

\acknowledgments We thank Giovanna Cottin and Yong Du for helpful discussion of the disappearing charged track searches. This work was supported in part by the
Australian Research Council. Feynman diagrams were drawn using
the TikZ-Feynman package~\cite{TikZFeynman}. The work of MJRM was supported in part under U.S. Department of Energy Contract DE-SC0011095 and National Natural Science Foundation of China grant number 19Z103010239. 

\bibliography{main}

\end{document}